\documentclass[aps, prb, twocolumn, showpacs, 10pt,superscriptaddress]{revtex4-1}
\usepackage{amsmath}
\usepackage{amsfonts}
\usepackage{amsbsy}
\usepackage{amssymb}
\usepackage{graphicx}
\usepackage{textcomp}
\usepackage[caption=false,singlelinecheck=false]{subfig}
\usepackage{xcolor}
\usepackage{mathrsfs}
\usepackage{mathtools}
\usepackage{bm}
\usepackage{gensymb}
\usepackage{braket,wasysym}
\usepackage[colorlinks,linkcolor=blue,citecolor=red,filecolor=magenta,urlcolor=cyan,breaklinks]{hyperref}
\usepackage[bottom]{footmisc}
\usepackage{verbatim}
\DeclareMathOperator{\sech}{sech}
\hypersetup{colorlinks=true, urlcolor=blue, citecolor=cyan, pdfborder={0 0 0}}
\usepackage{breakurl}
\usepackage{natbib}

\allowdisplaybreaks

\newcommand{\tcb}[1]{\textcolor{blue}{#1}}

\begin{document}
\title{Topological critical states and anomalous electronic\\ transmittance
 in one dimensional quasicrystals}

\author{Junmo Jeon}
\email{junmo1996@kaist.ac.kr}
\author{SungBin Lee}
\email{sungbin@kaist.ac.kr}

\affiliation{Korea Advanced Institute of Science and  Technology, Daejeon, South Korea}

\date{\today}
\begin{abstract}
Due to the absence of periodic length scale, electronic states and their topological properties in quasicrystals have been barely understood. Here, we focus on one dimensional quasicrystal and reveal that their electronic critical states are topologically robust. Based on tiling space cohomology, we exemplify the case of one dimensional aperiodic tilings especially Fibonacci quasicrystal and prove the existence of topological critical states at zero energy. Furthermore, we also show exotic electronic transmittance behavior near such topological critical states. Within the perturbative regime, we discuss lack of translational symmetries and presence of topological critical states lead to unconventional scaling behavior in transmittance. Considering both analytic analysis and numerics, electronic transmittance is computed in cases where the system is placed in air or is connected by semi-infinite periodic leads. Finally, we also discuss generalization of our analysis to other quasicrystals. 
Our findings open a new class of topological quantum states which solely exist in quasicrystals due to exotic tiling patterns in the absence of periodic length scale, and their anomalous electronic transport properties applicable to many experiments. 
\end{abstract}
\maketitle

\section{Introduction}
\label{sec:introduction}
Systems without periodicity are studied in various contexts such as  condensed matter physics, optics and mathematics \cite{sadun2008topology,suck2013quasicrystals,kawazoe2003structure,senechal1996quasicrystals,steinhardt1987physics,janot2012quasicrystals,kellendonk2015mathematics,divincenzo1999quasicrystals,macia2005role,yamamoto1996crystallography,vardeny2013optics,dal2003light,poon1992electronic}. In particular, quasicrystals which do not have periodic unit length scales and translational symmetries but show discrete diffraction measure, are mainly interested in condensed matter physics. For several decades since the discoveries of quasicrystals, many researchers are greatly interested in such non-periodic systems searching for new phases of matters with unconventional electronic and magnetic properties\cite{dal2003light,poon1992electronic,fukamichi1986magnetic,steinhardt2013quasicrystals,janot1996conductivity,hauser1986magnetic,stadnik2012physical,fuchs2018landau,rai2019proximity,kamiya2018discovery,deguchi2012quantum,Tanese2014FractalES,baboux2017measuring,rai2019proximity}. 
It has been studied that quasiperiodic system shows infinitely many gap structure in thermodynamic limit\cite{kalugin1986electron}. 
In addition, for some cases of quasicrystals, it has been investigated that there exist novel electronic states so called critical states where electron wave functions are neither completely localized nor extended\cite{mace2017critical,kohmoto1987critical}. 
Based on  several methods such as renormalization, transfer matrix and etc, above issues have been argued\cite{nishino1996corner,kreisel2016gabor,sadoc1993e8}. 

One of the most intriguing questions is whether such strange behavior addressed above is topologically robust and how it is directly related to anomalous physical phenomena so one can speculate new experimental signatures uniquely appear in quasicrystals.
However, due to absence of translational symmetries, various useful tools such as band theory and Bloch theorem which are generally used in periodic lattice systems, cannot be applicable to quasicrystals and it makes hard for us to understand physical properties in quasicrystals including energy spectrum, transport behavior and etc.
Instead, to answer the questions addressed above, one requires somewhat abstract approach that solely depending on tiling space of quasicrystal structures. 
Focusing on the pattern of the tiling itself, one can construct the metric space called tiling space. 
{In tiling space, abstract distance in between two distinct tilings say $T_1$, $T_2$ (they may or may not be similar tiling before deformation) where each tiling is composed of covering patterns of real space by using some polytopes (vertices, edges, faces etc), is given by the \textit{smallest translation} that makes two tilings identical through \textit{large region}. In detail, if two given tilings are congruent up to $\frac{1}{\epsilon}$ large region through translation on one tiling as amount of less than $\epsilon$, then we say that their distance is $\epsilon$ in abstract tiling space.}\cite{sadun2008topology,kellendonk2015mathematics}
Based on this kind of metric space for tilings itself, one can consider the \textit{pattern} dependent topologies.

In the context of such \textit{pattern} dependent topologies, two cohomology theories are mainly considered\footnote{The reason why we use cohomology here instead of homology or homotopy is because of the aperiodicity of quasicrystals, where the tiling spaces of them usually have infinitely many path connected components. Thus based on homology and homotopy, one cannot extract physically useful information}; one is \textit{Cech} cohomology and another is \textit{K-theory}\cite{kellendonk2015mathematics,sadun2008topology}. Roughly speaking, \textit{Cech} cohomology contains meaningful topological distinction between tilings and gives information related to diffraction measurement. \textit{K-theory}, on the other hand, which is the combination of cohomology and self-adjoint operator, gives the information of energy spectrum for pattern dependent hamiltonian. 
In terms of the K-theory, the elegant \textit{gap labeling theorem} has been well studied. Based on natural averaging trace map from K-theory abstract group to $\mathbb{R}$ space whose image is a group structure, it gives the information of integrated density of state (IDOS) of the spectral gaps.
This powerful theorem tells us by studying mathematical abstract structure of tiling space, one can (at least) identify exact positions of spectral gaps \cite{kellendonk2015mathematics,sadun2008topology,bellissard1992gap,kreisel2016gabor}.
In terms of the \textit{Cech} cohomology, it is more natural to concern the \textit{pattern equivariant cohomology} (a.k.a PE cohomology) even though mathematically \textit{Cech} cohomology and PE cohomology are equivalent. PE cohomology is nothing but the cohomology between chain complex that each cochain (you may think it as coloring of vertices, edges or faces with some \textit{pattern dependent} rule) is defined as containing (a part of) tiling \textit{pattern} information\cite{kellendonk2015mathematics}.
It is very robust under any kinds of local (or even nonlocal) perturbation which preserves tiling patterns. Thus, physical quantities which strongly depend on such PE cohomology or cochains, are considered as very robust concept\cite{kellendonk2015mathematics,kohmoto1987electronic,sadun2008topology,bellissard1992gap,kreisel2016gabor}. 
Such kind of discussion highly promoted understanding of quasicrystal or general aperiodic hamiltonian that lack of periodicity but depend on the pattern of the system.

In this paper, we apply such \textit{pattern} dependent topologies to the electronic system of the Fibonacci quasicrystal and discuss the topological critical state at zero energy and relevant electronic transports. 
Although the existence of critical states in quasicrystals has been known for a while, their topological aspects have not been understood\cite{mace2017critical,kohmoto1987critical}. 
Here, we show the critical state is a topologically robust quantity and give rise to anomalous transport properties. Topological perspectives provide significant understanding of critical states in quasicrystal which guarantees their robustness as long as their patterns are preserved in the presence of any (non-) local perturbations and even for strongly interacting electronic system.
In addition, based on both analytical and numerical studies, electronic transmittance and conductance near zero energy are also studied. 
Considering a simple tight binding limit with nearest neighbor hoppings, 
we discuss two situations where the system is placed in air or the system is connected with semi infinite periodic leads. 
It turns out that transmittance at zero energy shows a self-similar pattern along the system with nontrivial scaling behavior for weak quasi-periodic limit.
In addition, we also discover due to lack of periodic length scale, tiny control of total system size of quasicrystal enables huge change in transmittance near zero energy and such properties are also topologically protected, which can give many relevant experimental applications. 
Our theoretical approach based on PE cohomology and related transport properties can be generally applicable to other quasicrystal systems. 
Our work paves a way to uncover a new class of topological states which only exists in quasicrystals, distinct from periodic lattice system. Furthermore, it opens a way to understand how such topological electronic states are related to transport properties suggesting various future experiments. 

This paper is organized as following. First, we briefly review the  definition of Fibonacci quasicrystal and introduce some relevant previous works. Sec.\ref{sec:review} includes generating rule of Fibonacci quasicrystal, tight-binding Hamiltonian and zero energy state. In Sec.\ref{sec:Topology}, based on the PE cohomology group calculation, especially focusing on a Barge-Diamond complex, we prove that the zero energy critical state is indeed topologically robust. In Sec.\ref{sec:Transmittance}, we analytically study electronic transmittance near topological critical state and provide the explicit expression of transmittance within perturbative regime. Supporting numerical results are also presented in this section with discussions. Summary and some remarks in our results are given in Sec.\ref{sec:results}. 

\section{Review : Generation of Fibonacci quasicrystal and zero energy critical state}
\label{sec:review}
In this section, we introduce the definition of Fibonacci quasicrystal and briefly review the zero energy critical states.
The Fibonacci quasicrystal can be generated using substitution method.
It is composed of two prototiles which are called $L$(=Long) and $S$(=Short). Starting with prototile $L$, we successively apply substitution in the following ways\cite{levine1986quasicrystals,gumbs1988dynamical},
\begin{align}
\label{eq:1}
&\sigma=\begin{cases}L \to LS \\ S \to L.
\end{cases}
\end{align}
By choosing the basis as a prototile itself, the substitution can be also rewritten as
\begin{align}
\label{eq:2}
&\begin{pmatrix} LS  \\ L \end{pmatrix}=\begin{pmatrix} 1 & 1 \\ 1 & 0 \end{pmatrix}\begin{pmatrix} L \\ S \end{pmatrix}
\end{align}
Here, the above $2\times 2$ matrix is called substitution matrix. It contains all the special characteristics of aperiodicity such as self-similarity of the tiling itself. Since the matrix characteristic equation is given as $\lambda^2-\lambda-1=0$ whose solution is $\tau$ and $1/\tau$ where $\tau$ is a golden ratio, it satisfies a Pisot condition i.e., only one of the eigenvalues, $\tau$, is larger than 1. This is a necessary condition of discrete diffraction measure\cite{baake_grimm_2013}. 
Its PE cohomology (or equivalently \textit{Cech} cohomology) is known as $\mathbb{Z}^2$ and this classifies possible deformation  which preserves diffraction pattern\cite{sadun2008topology}. 
On the other hand, the gap labeling group is known as $\mathbb{Z}\oplus\tau\mathbb{Z}$ which can be obtained by a trace map 
and this is directly related to the energy spectrum with infinitely many spectral gaps whose IDOS belong to $\mathbb{Z}\oplus\tau\mathbb{Z}$\cite{sadun2008topology}. 


We consider the spinless tight-binding model with the nearest-neighbor hopping,
\begin{align}
\label{eq:3}
&H=\sum_{\left\langle{i,j}\right\rangle}(\ket{i}t_{i,j}\bra{j}+h.c.)+\sum_{i}\ket{i}\epsilon_{i}\bra{i}.
\end{align}
Here $i$ and $j$ are atomic site positions. $t_{i,j}$  and $\epsilon_{i}$ are hopping constant between neighboring sites $i$ and $j$ and on site energy at site $i$ respectively. There are two ways to impose the quasi-periodicity; one is to put quasi-periodic structure in the hopping term $t_{ij}$ and another is to put it in on site energy. Physically, the former corresponds to control the distance between each atomic site, the latter corresponds to control atomic kinds. 
We note that in considering transfer matrix and transmittance, the latter case is not much interesting. When quasi-periodicity is given in on site energies, simple addition of entire onsite energies or their \textit{arbitrary} combination products appear in the transfer matrix. Then it is meaningless because simple addition is commutative and this corresponds to the same as rearranging prototiles like  periodic case in some sense.
In the former case, however, quasi-periodicity character is fully considered  in the transfer matrix and gives rise to anomalous transmittance as we will discuss later. Hence, we focus on the former case only and without loss of generality we set $\epsilon_{i}=0$. 

Due to the sublattice symmetry or particle-hole symmetry, the energy spectrum of Eq.\eqref{eq:3} contains both $E$ and $-E$. Since the gap labeling group of the Fibonacci quasicrystal tiling space is $\mathbb{Z}\oplus\tau\mathbb{Z}$ and does not contain 1/2, it tells us that zero energy electron state exists which is doubly degenerate. 
It is worth to note that silver mean, cantor set and binary nonpisot tilings also belong to this kind of tilings, whereas 
the gap labeling groups of Thue-Morse and paper folding tilings contain 1/2. In other words, if we consider tight binding Hamiltonian of Thue-Morse tiling, for instance, there is no zero energy state. Since gap labeling group itself is a topological quantity, the existence of the gap in IDOS is also topologically protected\cite{sadun2008topology}.

Now let's consider zero energy states. Explicitly, the Schr\"{o}dinger equation is written as following.
\begin{align}
\label{eq:4}
&t_{n+1}\psi(n+2)+t_{n}\psi(n)=0,
\end{align}
where $\psi(n)$ is the electron wave function at $n$-th site and $t_{n}$ is hopping magnitude from $n$-th site to $(n+1)$-th site.
Particularly for the Fibonacci quasicrystal case (in principle, it can be applicable for any arbitrary systems composed of two prototiles), one can define the systematic parameter $\rho\equiv t_{L}/t_{S}$, where $t_{L}$ and $t_{S}$ are hopping magnitudes for long tiling and short tiling respectively. Then, $\psi({2(n+1)})$ is represented as,
\begin{align}
\label{eq:5}
&\psi(2(n+1))=-\rho^{a(2n\to2(n+1))}\psi(2n).
\end{align}
Here, $a(2n\to2(n+1))$ is a function of the hopping constant arrangement between $2n$-th site and $2(n+1)$-th site. Explicitly, $a(LL)=0, a(LS)=1, a(SL)=-1$ and note that there is no $SS$ supertile\cite{senechal1996quasicrystals}. By applying it inductively, one gets
\begin{align}
\label{eq:7}
\psi(2n)=(-1)^n\rho^{h(n)}\psi(0)
\end{align}
where $h(n)=\sum_{0\le i<n}a(2i\to2(i+1))$. Thus, it leads to
\begin{align}
\label{eq:8}
&\frac{\psi(2n)}{\psi(0)}=(-1)^n\exp(\kappa h(n)),
\end{align}
with $\kappa=\ln{\rho}$. The square of Eq.\eqref{eq:8} is nothing but one of the two eigenvalues for the transfer matrix product $MM^\dagger$. Note that its eigenvalues are inverse pairs. 
In above procedure, selection of even sites in Eq.\eqref{eq:5} gives one of the degenerate states. Another degenerate state can be obtained by selecting only odd sites. The only difference in that case is to replace $h(n)$ into $f(n)=\sum_{0\le i<n}a(2i+1\to2i+3)$. 
Their critical characters and fractal dimensions are already well studied\cite{mace2017critical}.

\section{Topology of supertiling space and protected critical state}
\label{sec:Topology}

In this section, we show that the functions $h(n)$ and $f(n)$ introduced in Sec.\ref{sec:review} are essentially topological ones, thus claim that they are robust under any kinds of pattern equivalent transformations including simple translations and etc. Based on our argument, we show that zero energy critical state is indeed a topologically protected phase. In order to show such property, we adopt pattern dependent topologies and derive PE cohomology group of supertile in particular. Furthermore, as a consequence of topological property, we also address unique logarithmic scaling behavior of $h(n)$ and $f(n)$ in the Fibonacci quasicrystal and compare them with other aperiodic systems.

\begin{figure}[!h]
\centering
\subfloat[]{\includegraphics[width=0.3 \textwidth]{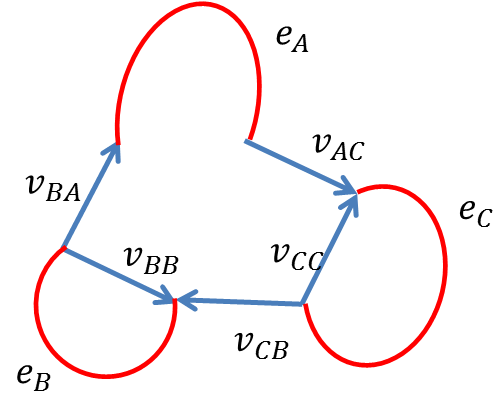}
\label{fig:bargediamonda}}
\hfill
\subfloat[]{\includegraphics[width=0.5 \textwidth]{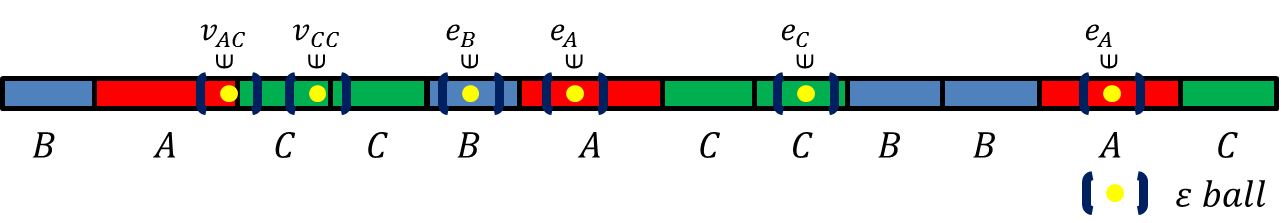}
\label{fig:bargediamondb}}
\caption{\label{fig:bargediamond} (a) Barge-Diamond complex of the Fibonacci quasicrystal supertiling. Here, the edges $e_{A,B,C}$ are equivalence class of real space position whose neighborhood is completely included in A,B,C supertile respectively. The vertex flips $v_{ij}$ where $i,j=A,B,C$ represent whose neighborhood experience edge flips from $i$ to $j$. Examples of equivalence class with some $\epsilon$ ball is shown in (b). See the main text for more details.
}
\end{figure}

We first consider supertiles with length 2 in the Fibonacci tiling. There exist three types of colored supertiles $LL, LS, SL$ and let's call them $A,B,C$ respectively. (Note that we don't have a supertile of $SS$ type.)
As an example, this allows us to rewrite the Fibonacci tiling of the first few words as, 
\begin{align}
\label{eq:9}
&LSLLSLSL \to BACC.
\end{align}
By explicitly applying substitution in Eq.\eqref{eq:1}, we have a new substitution matrix for supertiling in the basis of ${A,B,C}$,
\begin{align}
\label{eq:10}
&S=\begin{pmatrix} 1 & 1 & 1 \\ 2 & 1 & 2 \\ 2 & 2 & 1 \end{pmatrix}.
\end{align}
Now let's consider the PE cohomology of supertiling. In order to compute cohomology group, we construct the Barge-Diamond complex as shown in Fig.\ref{fig:bargediamonda}. 
The Barge-Diamond complex is mainly composed of the two types of equivalence classes, edges $e_i$ and vertex flips $v_{ij}$ ($i,j=A,B,C$). As an example, Fig.\ref{fig:bargediamondb} shows the Fibonacci quasicrystal supertiling $BACCBACCBBAC$ along the line which can be easily gotten after a couple of substitution. Here, the edge $e_i$ is defined for within a given $\epsilon$ ball regime when neighborhood of the point is completely included in $i$ supertile, while the vertex flip $v_{ij}$ is defined for the case where neighborhood experience the change of supertile from $i$ to $j$. Each point on the line belongs to an equivalence class, if their neighborhood pattern is equivalent within small range given as $\epsilon$ ball. 
Since couple of successive words are forbidden in the system, such as $LLL$ and $SS$, there are only 5 kinds of vertex flips between two supertiles $v_{BA}, v_{AC}, v_{BB}, v_{CB}, v_{CC}$.
(A proof by contradiction --- Assuming successive words $LLL$ or $SS$ are appeared, apply inverse substitution so that it cannot be returned into a single prototile, $S$.)
 In addition, rule of the Fibonacci tiling gives unique orientations of vertex flip between $A,C$ and $A,B$ and it determines the arrow directions of $v_{BA}$ and $v_{AC}$ shown in Fig.\ref{fig:bargediamonda}. It indicates that a vertex flip from $C$ supertile to $A$ supertile is forbidden but a flip from $A$ to $C$ is allowed. Similarly, a vertex flip from $B$ to $A$ is allowed but a flip from $A$ to $B$ is forbidden. 
Others are given as compatibly. 
This complex completely contains the pattern information of the Fibonacci quasicrystal supertiling and
we also note that the collection of $v_{ij}$ has no additional loop. 

Now, we discuss the Barge-Diamond complex in the thermodynamic limit. 
By keep applying the substitution matrix  Eq.\eqref{eq:10}, one can easily read off that only two types of vertex flips $v_{BB}$ and $v_{CB}$ survive. For instance, if we apply a single substitution on an equivalence class of point $p$ on $v_{AC}$, then it is mapped into $v_{BB}$ since the tiling after a single substitution becomes $BACCB(p)BACB$. 
Similarly, the vertex flips $v_{BA}$ and $v_{CC}$ are mapped into $v_{CB}$ and $v_{BB}$ respectively, whereas $v_{BB}$ and $v_{CB}$ are mapped into $v_{CB}$ and $v_{BB}$ respectively. 
Thus, only $v_{BB}$ and $v_{CB}$ survive and this leads  the eventual range of a Barge-Diamond subcomplex to be a single connected component. Hence, the PE cohomology group of the Fibonacci supertiling space $T$ can be derived by (reduced) cohomology exact sequence for the pair $(\Xi,\Xi_0)$ consisting of inverse limit of a complex $\Xi$ and a subcomplex $\Xi_0$\cite{sadun2008topology}. Eq.\eqref{eq:longexact} shows a long exact sequence of cohomology group.  
%
\begin{align}
\label{eq:longexact}
& 0\to \check{H}^0(\Xi,\Xi_0)\to \check{H}^0(\Xi)\to \check{H}^0(\Xi_0)\nonumber \\
&\to \check{H}^1(\Xi,\Xi_0)\to \check{H}^1(\Xi) \to \check{H}^1(\Xi_0) \to 0.
\end{align}
Here, arrow implies that (cohomology) group homomorphism. ``Exact" means that image of each homomorphism is equal to the kernel of the next and $0$ stands for a trivial group. $\check{H}^i(\Xi)$ stands for the $i$-th PE reduced cohomology group and $\check{H}^i(\Xi,\Xi_0)$ is the $i$-th relative reduced cohomology group. In our case, since $\Xi,\Xi_0$ and quotient of $\Xi$ by vertex flips are connected clearly, all zeroth reduced cohomologies are vanished resulting in $\check{H}^0(\Xi, \Xi_0)=\check{H}^0(\Xi)=\check{H}^0(\Xi_0)=0$.
Here, the quotient of the Barge-Diamond complex by vertex flip subcomplex, is the wedge of three circles corresponding to three supertiles. Thus, in our case of the Fibonacci quasicrystal, $\check{H}^1(\Xi,\Xi_0)$, which is defined as the direct limit of $S^T$ (transpose of the substitution matrix given in Eq\eqref{eq:10})  i.e., a quotient space with $S^T$ as a quotient map,  is $\mathbb{Z}^3$. (Note that the eigenvalues of $S^T$ are irrational and $-1$, hence there is no effect due to the quotient on each $\mathbb{Z}$.)
  Now it can be shown from substitution rules that the eventual range of subcomplex of the substitution map is a single connected component and contractible (because there is no loop). Hence, we conclude that $H^1(T)=\mathbb{Z}^3$ for the Fibonacci supertiling where $H^i(T)$ is the $i$-th cohomology group (not reduced one) of Fibonacci supertiling space. Physically, it means that when we consider thermodynamic limit of the Fibonacci tiling, there are $\mathbb{Z}^3$ kinds of nontrivial topologically protected quantities. If one can find the connection between one of them and it's relevant physical quantities, then it implies that such physical quantity should be topologically protected. 
In the next section, we will show that it is indeed the case of transmittance near the zero energy in the Fibonacci quasicrystal. The zeroth cohomology is trivial from the definition of the reduced cohomology.
Thus, combining both the zeroth and the first PE cohomology groups of the Fibonacci supertiling space $T$, one obtains,
\begin{align}
\label{eq:cohomology}
&H^0(T)=\mathbb{Z}, H^1(T)=\mathbb{Z}^3.
\end{align}

\begin{figure}[!h]
\centering
\includegraphics[width=0.5 \textwidth]{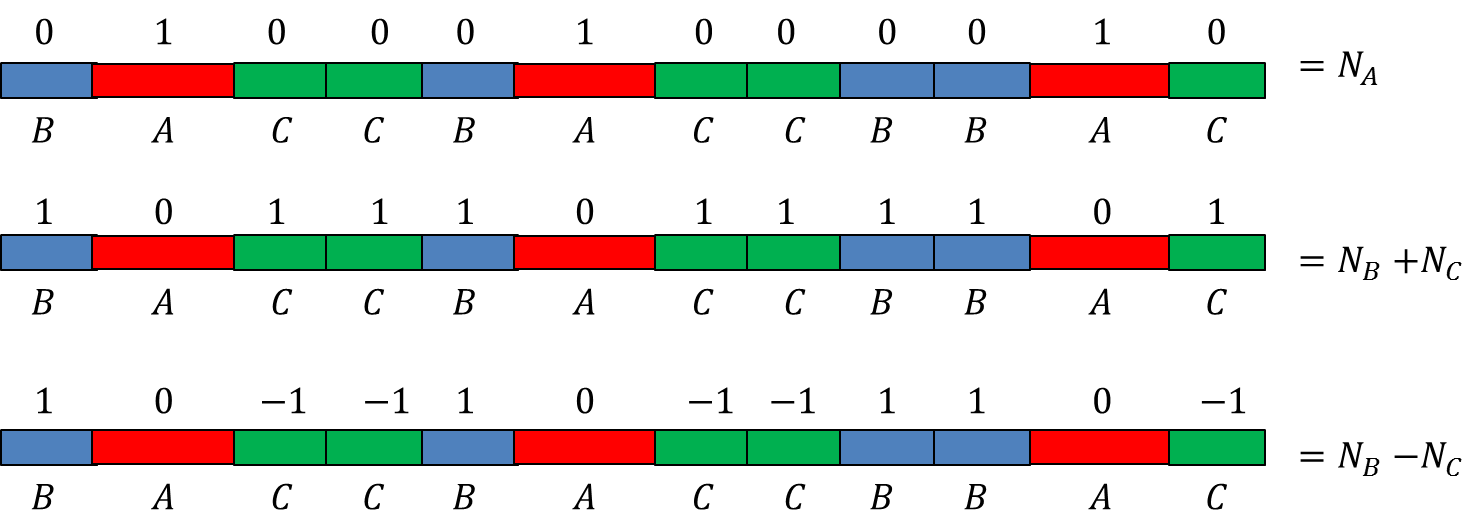}
\caption{\label{fig:generators}Generators of the first cohomology group for the Fibonacci supertilling space, $H^1(T)$. Each generator comes from the simple counting number of each supertiling, $N_A,N_B,N_C$.} 
\end{figure}

Explicitly, one way to construct the generators of $H^1(T)$ is given in Fig.\ref{fig:generators}. In Fig.\ref{fig:generators}, $N_{A,B,C}$ stands for total number of $A,B,C$ supertile of the system respectively. 
Clearly from the definition, $h(n)$ introduced in Eq.\eqref{eq:7} is nothing but (integrated) 1-cochain function of the Fibonacci supertiling with the element $N_B-N_C$ and it can be obtained from the third generator shown in Fig.\ref{fig:generators}. 
Having known that the element $N_B-N_c$ is a topological quantity, let's now prove by contradiction, that it is a \textit{nontrivial} element.
Let's assume that $N_B-N_C$ is trivial i.e., one of the elements in the image of coboundary map called $\delta$. Then there exists a pattern equivariant 0-cochain (vertex flips) map, say $\beta$, such that $N_B-N_C=\delta \beta$. Here, pattern equivariant means that at least if two vertices, say $x,y$ are in the same vertex flip, then $\beta(x)=\beta(y)$. However, by simple checking along Fibonacci tiling, one can easily show that it is impossible to find such $\beta$. 
For instance, in the third figure in Fig.\ref{fig:generators}, if we let $x,y$ be the first and the second vertices for $v_{BA}$, then $\beta(x)=\beta(y)+1\neq \beta(y)$. Thus, by contradiction, it implies that $N_B-N_C$ is a \textit{nontrivial} element in cohomology group. 
Hence, we can conclude that $h(n)$ is indeed a nontrivial topologically invariant quantity.

Compared to $h(n)$, $f(n)$ is just translationally shifted one from even sites to odd sites. Hence, based on the definitions of $f(n)$ and $h(n)$, one can clearly read off,
\begin{align}
\label{eq:fh}
&f(n)+h(n)=0,1.
\end{align}
Here, the difference between $0$ and $1$ comes from end boundary condition; If the end boundary of the supertiling is either $A$ or $B$, then it is zero. If the end boundary is $C$ supertile, then it becomes 1. 
In terms of the original Fibonacci tiling not on the basis of supertiling, if the type of $(2n+1)$-th tile is $L$, then it is zero, but if it is $S$, there is an offset by 1. Thus, one can extract $n$, such that there is an offset between $f(n)$ and $h(n)$ by using the position of the $S$ prototile. This property becomes significantly important for the transmittance of the Fibonacci quasicrystal connected with periodic leads, which we will discuss in the next section. 
Eq.\eqref{eq:fh} indicates that $h(n)$ and $f(n)$ both have common scaling behavior up to overall sign. In addition, $f(n)$ itself is also a topologically robust quantity same as $h(n)$, thus it does not change under any kinds of pattern preserving perturbations or local operations. But, we note that they are not cohomologous because $f(n)-h(n)=-2h(n)+(0 , \pm1)$ is not coboundary.
It implies that $h(n)$ and $f(n)$ may explain independent physical phenomenon which will be discussed in Sec.\ref{sec:Transmittance}.

Topological robustness of $h(n)$ (similarly for $f(n)$) is used to classify the aperiodic systems. It is because the scaling characters of extended, critical and localized states entirely depend on the scaling behavior of $h(n)$. 
In particular, scaling behavior of the critical state is given as 
\begin{align}
\label{eq:19}
&h(L),f(L)\thicksim\sqrt{\ln{\left( \frac{L}{L_0} \right)}},
\end{align}
where $L$ is the total system size in terms of the tiling unit 
and $L_0$ is the length of the short prototile which is set to make dimensionless quantity in the logarithm. Detailed derivation of Eq.\eqref{eq:19} is given in Supplementary Materials Sec.\ref{subsec:Scaling behavior of h(n)}.
As a consequence of topological robustness, existence of the critical state is also topologically protected and scaling behavior of the critical state cannot be changed from any local or pattern equivariant perturbations.

In order to further understand topological robustness of the critical state, we consider three more examples and compare them with the Fibonacci quasicrystal case; silver mean, Cantor set and binary non-Pisot system. All of them belong to aperiodic tilings which are well studied in several literatures\cite{kellendonk2015mathematics,baake2013aperiodic,walter2009crystallography}. Similar to the Fibonacci quasicrystal, they are all composed by two prototiles, say $A$ and $B$ and each of them are defined by the following substitution rules. 
\begin{eqnarray}
\label{eq:three-tilings}
\sigma_{\text{SM}}~&=&\begin{cases}A \to BAA \\ B \to A  \end{cases} \\
\sigma_{\text{CS}}~&=&\begin{cases}A \to ABA \\ B \to BBB  \end{cases} \nonumber \\
\sigma_{\text{B-NP}}&=&\begin{cases}A \to AB \\ B \to AAAAA  \end{cases} \nonumber
\end{eqnarray}
Here, $\sigma_{\text{SM,CS,B-NP}}$ represent substitution maps for silver mean, Cantor set and binary non-Pisot systems respectively. Based on the substitution rules given in Eq.\eqref{eq:three-tilings}, one can easily compute the eigenvalues of substitution matrices and find that only the silver mean tiling satisfies a Pisot condition like the case of Fibonacci quasicrystal. On the other hand, other two tilings are well known as non pisot substitution tiling, i.e., random tilings which show absolute continuous diffraction measure\cite{kellendonk2015mathematics}. 
The gap labeling groups of these three tilings are already well known as $\mathbb{Z}+(\sqrt{2}+1)\mathbb{Z}$ and $\frac{k}{3^N}$, $\frac{\mathbb{Z}+\omega \mathbb{Z}}{5^N}$ respectively where $k,N$ are integers and $\omega=\frac{\sqrt{26}+1}{2}$\cite{kellendonk2015mathematics}. 
Note that as we have already mentioned in Sec.\ref{sec:review}, such gap labeling groups imply that there exist zero energy states for those tilings.

\begin{figure}[!h]
\centering
\includegraphics[width=0.5 \textwidth]{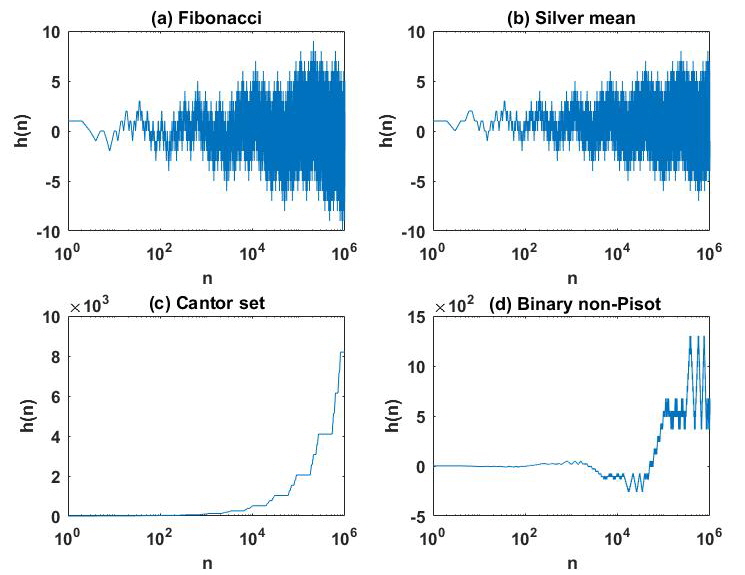}
\caption{\label{fig:2} $h(n)$ as a function of $n$ up to $n=10^6$ (log scale) for various tilings in 1D. (top left) Fibonacci (top right) silver mean (bottom left) Canotr set (bottom right) binary nonpisot. Based on Eq.\eqref{eq:8}, growth speed of $h(n)$ as a function of $n$ implies the characteristic of zero energy eigenstate, either critical state (Fibonacci, silver mean) or localized state (Cantor set, binary non-Pisot). See the main text for more details. }
\end{figure}

Now we discuss the zero energy state for each tiling case. Their features are totally distinguishable between the two groups, Fibonacci and silver-mean tilings (Pisot quasicrystals) \textit{vs} Cantor set and binary non-Pisot systems. In Fig.\ref{fig:2}, behavior of $h(n)$ as a function of $n$ is shown for each tiling. 
By substituting the scaling behavior of $h(L)$ into Eq.\eqref{eq:8}, one can easily determine characteristics of zero energy state whether extended, localized or critical. Fig.\ref{fig:2} shows the growth speed of $h(n)$ as a function of $n$ for each case. As seen in Fig.\ref{fig:2}, the growth speed of $h(n)$ indicates that the zero energy states of both Fibonacci and silver mean tilings are critical, whereas the system contains localized zero energy states for the case of a Cantor set and a binary non-Pisot system. 
Importantly, one can observe that the behavior of $h(n)$ for Fibonacci and silver mean cases are almost identical despite their distinct tilings. 
These behavior is guaranteed from the first cohomology groups of supertilings. Both Fibonacci and silver mean tilings have the same first cohomology group $\mathbb{Z}^3$ and $h(n)$ in both cases is described by topologically nontrivial element $\mathbb{Z}$ for supertilings. (Detailed derivation of cohomology group for silver mean tiling is given in Supplementary Materials Sec.\ref{subsec:silvermean}.) 
On the other hand, both a Cantor set and a binary non-Pisot system have distinct first cohomology groups;  $\mathbb{Z}[1/3]\oplus\mathbb{Z}[1/2]^2$ for a Cantor set  and $\mathbb{Z}[1/5] \oplus \mathbb{Z}^2$  for a binary non-Pisot system. In particular, $h(n)$ for Cantor tiling is described by one element of $\mathbb{Z}[1/2]$, whereas, $h(n)$ for a binary non-Pisot system is described by one  element of $\mathbb{Z}[1/5]$. (See Supplementary Materials Sec.\ref{subsec:silvermean})

Note that $\mathbb{Z}[1/2]$ and $\mathbb{Z}[1/5]$ implies the cyclic inner structure of the whole group. As an example, $\mathbb{Z}[1/2]$ includes all the elements $\left\{ \frac{1}{2},1 \right\}, \left\{ \frac{1}{4},\frac{2}{4},\frac{3}{4},1 \right\},\cdots$. That means our nontrivial elements appear in infinitely many cyclic structures and all of them are indistinguishable. For example, $1=\frac{1}{2}+\frac{1}{2}=\frac{1}{4}+\frac{1}{4}+\frac{1}{4}+\frac{1}{4}=\cdots$. Thus, in such case, an infinite number of inner structures of cochains exist, which are composed of fractional elements such as $1/2$ or $1/4$ as shown in above example.  
 It means that a cochain function itself (before integration) is repetitive with a \textit{finite nontrivial value}, say $\gamma$ which originates from fractional elements. Then, an integrated cochain function, $h(n)$, becomes repetitive sum of $\gamma$. Hence, $h(n)$ is growing fast with the system size. On the other hand, in the Fibonacci or silver mean tilings, such nontrivial inner structures of cochain are absent and the speed of growth of $h(n)$ is suppressed\footnote{Junmo Jeon and SungBin Lee, in preparation.}.
It yields localization behavior of Cantor and binary non-Pisot systems, distinct from Fibonacci and silver mean tilings.
It is worth to note that although Cantor and binary non-Pisot systems both exhibit localized zero energy states, their localization behavior is quite different. 
In case of a Cantor set , $h(n)$ shows a stair shape which is rapidly and monotonically increasing\cite{kellendonk2015mathematics}. 
On the other hand, in a binary non-Pisot system, it shows a steeply oscillating behavior. It implies that even though both cases have localized zero energy states, they have different localization behavior. For the case of a Cantor set, it has a finite localization length which is roughly given by a range of constant $h(n)$. In contrast, for a binary non-Pisot system, the zero energy state is strongly localized.

{In this section, we have computed explicit cohomology group of the Fibonacci supertiling space using the Barge-Diamond complex. By studying the first  cohomology group $H^1(T)=\mathbb{Z}^3$, we conclude that $h(n)$ and $f(n)$ are directly described by the nontrivial elements of $\mathbb{Z}$ sector, thus the zero energy critical state is indeed a topologically protected nontrivial quantity. 
We have also shown it's topological robustness comparing with other aperiodic tilings such as silver-mean, Cantor set and binary non-Pisot systems. 
Despite distinct tiling patterns especially for Fibonacci and silver-mean cases, the tilings for equivalent supertiling cohomology group share the same nontrivial critical behavior. 
Even though we exemplify the Fibonacci quasicrystal case, our approach based on cohomology group of supertilings can be applicable to any kinds of tilings and this method can be generally used to classify characters of critical states in several aperiodic tilings.
Moreover, this robust topological quantity is closely related to the important transport phenomena which we will discuss in the following section. }

\section{Transmittance}
\label{sec:Transmittance}

\begin{figure}[!h]
\centering
\includegraphics[width=0.5 \textwidth]{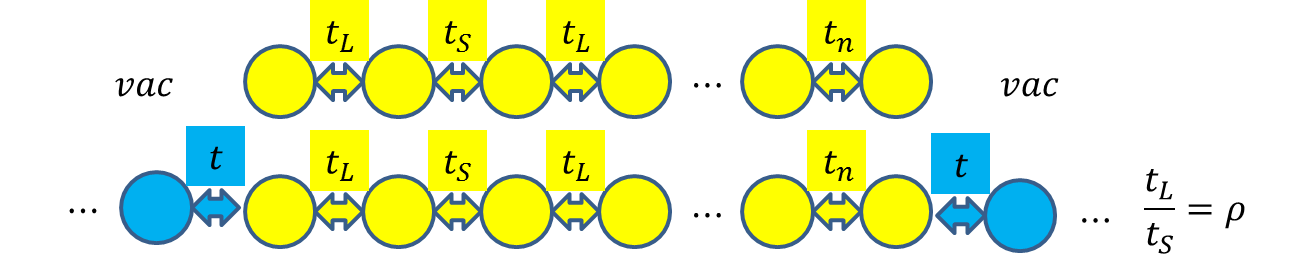}
\caption{\label{fig:system} Schematic picture for two distinct boundary conditions. Top figure represents the situation where the Fibonacci quasicrystal (colored in yellow) is placed in air. Bottom figure represents the case connected with two periodic leads (colored in blue). } 
\end{figure}

Based on the topological critical state obtained in Sec.\ref{sec:Topology}, we now discuss anomalous transport properties of the Fibonacci quasicrystal. 
Since transport itself strongly depends on energy eigenstates of the system, given that the corresponding state is topological, the transport properties should be also topologically protected. In other words, topological electronic state automatically guarantees some robust quantity of electronic transport as well and such robustness is valid even with adding a small amount of energy. 
We take into account electronic transmittance for two different boundary conditions; (1) One dimensional system  of the Fibonacci quasicrystal is placed in air.  (2) The system is connected by semi-infinite periodic leads. Fig.\ref{fig:system} shows the schematic picture of these two cases. Considering topological critical state at zero energy, we focus on the transmittance near zero energy within perturbative regime.  
Both analytical and numerical analysis of electron transmittance and conductance are given for the case of the Fibonacci quasicrystals. We also discuss our results compared to other aperiodic tilings.

Before discussing the quasicrystal case, let's briefly argue electronic transmittance of the periodic cases. 
When the system is perfectly periodic, the results are already well known. For instance, when a single prototile periodic system is placed in air, perfect transmittance is expected due to the Bloch theorem for any allowed energy. For bi-prototile case, however, there is no transmittance at and nearby zero energy which is understood in various points of view especially in terms of the band structure\cite{Simon:1581455}. In the periodic case, therefore, the zero temperature conductance is quantized by $\frac{2e^2}{h}$. 
However, such argument is \textit{not} valid anymore in quasicrystals and we show how it behaves differently compared to the case of periodic system. 

\begin{figure}[!h]
\centering
\includegraphics[width=0.5 \textwidth]{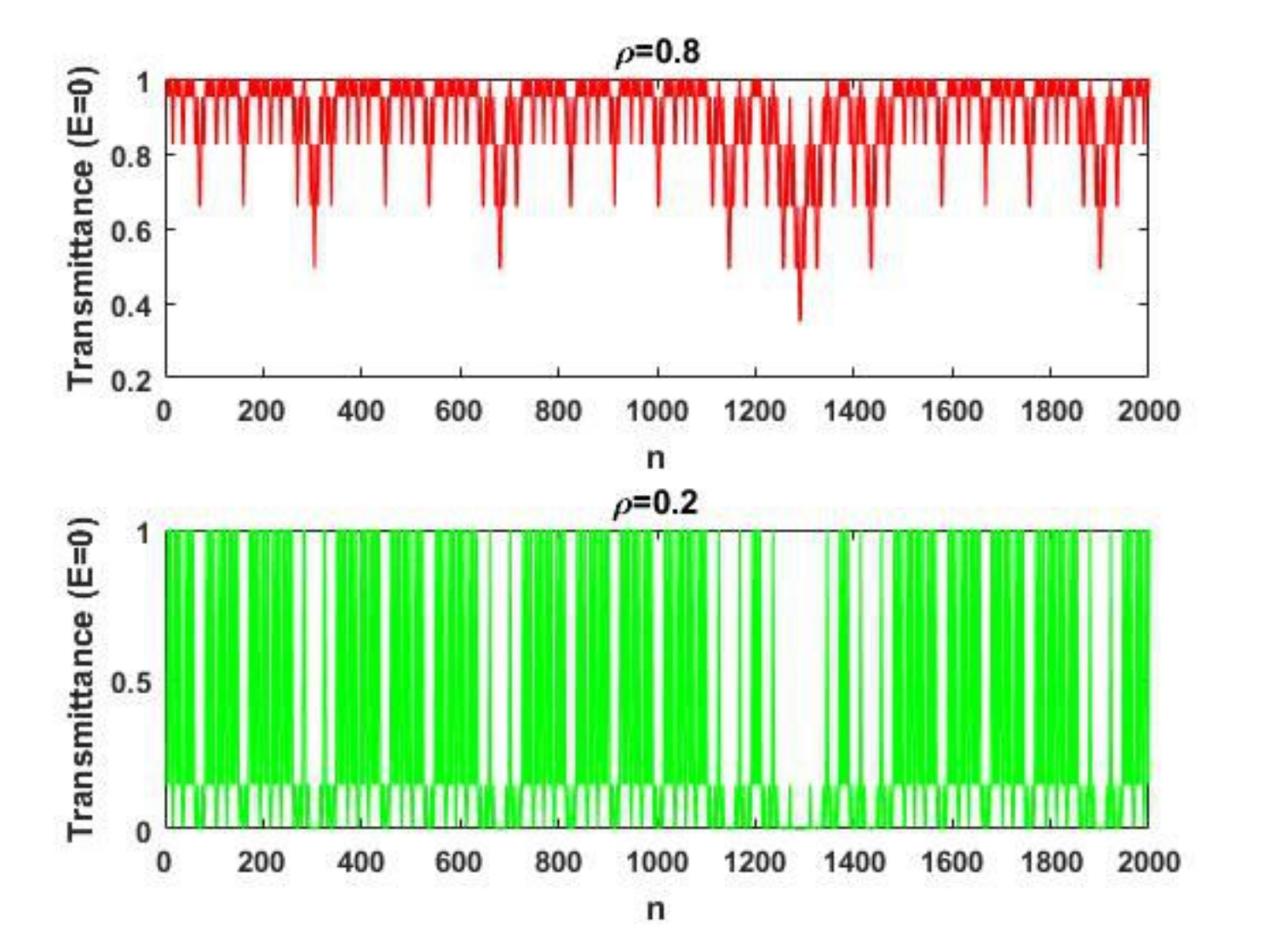}
\caption{ Zero energy transmittance $T_{n=2k}$ given in Eq.\eqref{eq:20} of the Fibonacci quasicrystal for $\rho~(=t_L/t_S)=0.8$ (top) and  $\rho=0.2$ (bottom). 
Presence of many sites $n$ having perfect transmittance ($T_{n}\!=\!1$) indicates a special character of the critical state and such non-trivial sites $n$ do not depend on the strength of quasi-periodicity. 
}
\label{fig:3}
\end{figure}

We first consider transmittance of the zero energy state. Both air boundary or semi-infinite periodic lead boundary share the similar behavior in this case, so without loss of generality we focus on the former case. (The only difference between the former and the latter cases may be interchanging between $h(n)$ and $f(n)$ which will be discussed in \ref{subsec:Transmittance with two conducting leads} in detail.)
In general, transmittance $T_{n=2k}=\sech^{2}{(x_n)}$ is defined for $e^{x_n} \equiv \left|\frac{\psi(n)}{\psi(0)}\right|$\cite{beenakker1997random}. Thus, transmittance in terms of $\kappa$ and $h(k)$ from Eq\eqref{eq:8} is represented as,
\begin{align}
\label{eq:20}
&T_{n=2k}=\frac{1}{\cosh^2{(\kappa h(k))}}.
\end{align}
This formula comes from one of the eigenvalue of product of transfer matrix $MM^\dagger$, $e^{2x_n}$, whose another eigenvalue is always it's inverse $e^{-2x_n}$. 
  Depending on the strength of quasi-periodicity $\rho=t_L/t_S=0.8$ and $0.2$, Fig.\ref{fig:3} shows the zero energy transmittance of the Fibonacci quasicrystal up to $n=2000$ sites.

In the thermodynamic limit with infinite total length of the Fibonacci tiling, the number of $B$-type supertile becomes equivalent to  that of $C$-type supertile,  $N_B=N_C$.\cite{mace2017critical} It implies that there are many nontrivial $n$ such that $T_{n=2k}=1$ with perfect transmittance as shown in Fig.\ref{fig:3}. This is the special character of the critical states, which does appear neither in localized states nor in extended states. Furthermore, from Eq.\eqref{eq:20}, it is clear that specific values of such $n$ do \textit{not} depend on the strength of quasi-periodicity. As seen in Fig.\ref{fig:3}, patterns of the zero energy transmittance as a function of sites also shows self similarity which originates from the self-similar behavior of $h(k)$. 

Let's discuss the arithmetic average of transmittance to derive the scaling behavior as a function of system size,
\begin{align}
\label{offset}
&\bar{T}=\frac{1}{n}\sum_{i\le n}T_{i}.
\end{align}
In general, there is transmittance decay in the critical states unlike the extended states for the periodic case. Such decay can be analytically computed from the behavior of $h(k)^2$. (See Supplementary Materials Sec.\ref{subsec:Scaling behavior of h(n)}). For weak quasi-periodicity limit i.e. $\rho = e^\kappa \approx 1$, the leading order of the Taylor series expansion in terms of $\kappa h(k)$ in Eq.\eqref{offset} results in $\bar{T} \approx 1-\frac{\kappa^2}{n}\sum_{i}h(i)^2$.  In the thermodynamic limit, we finally get the scaling behavior up to the leading order as following.
\begin{align}
\label{offset2}
&1-\bar{T}\approx\frac{\ln(\lambda(\kappa)/\lambda(0))}{\ln \tau^3}\ln\left(\frac{L}{L_0}\right),
\end{align}
where $L$ is the total system size in terms of the tiling unit and $L_0$ is the length of the short prototile which is again set to make dimensionless quantity in the logarithm. Here, $\lambda$ is defined as,
\begin{align}
\label{lambda}
&\lambda(\beta)=\left( \frac{(1+e^\beta)^2+\sqrt{(1+e^\beta)^4+4e^{2\beta}}}{2e^{\beta}} \right)^2.
\end{align}
See Supplementary Materials Sec. \ref{subsec:Scaling behavior of h(n)} for detailed information of $\lambda(\beta)$.
 This logarithmic scaling behavior can be obtained for small $\left| \kappa \right |$ which is the limit of weak quasi-periodicity. Note that for perfect periodic case, the right hand side of Eq.\eqref{offset2} is clearly zero and there is no deviation from the perfect transmittance ($\bar{T} = 1$). 
In the limit of strong quasi-peirodicity with $\left |\kappa \right | \gg 1$, however, Eq.\eqref{offset2} does not hold anymore and there exist no closed form in terms of the total system size $L$. In this case, the average transmittance $\bar{T}$ becomes zero much faster as a function of $L$, compared to the case of weak quasi-periodic limit.

Important remark of Eq.\eqref{offset2} is the following. Since Eq.\eqref{offset2} is originated from the topological quantity, $h(k)$, this scaling behavior of average transmittance $\bar{T}$ is also topologically protected.
One may argue the scaling behavior changes for different tilings. However, as we have already mentioned in Sec.\ref{sec:Topology}, the tilings which share the same cohomology group (like Fibonacci and silver mean tilings) show very similar behavior. Therefore, classification of aperiodic tilings can be also performed by measuring how average transmittance $\bar{T}$ in the limit of weak quasi-periodicity decays as the total system size increases.

\subsection{Transmittance with air boundary near zero energy}
\label{subsec:Transmittance with air boundary}

Now we take into account the case where the Fibonacci quasicrystal is placed in air and calculate the transmittance \textit{near} zero energy. (Note that the following approach is generally applicable to any systems.) In this case, the boundary condition is the Dirichlet condition, since there is no hopping outside of the system $\psi(-1)$=$0$ and we consider generation of plane wave at zeroth position ($i=0$). 
Let's consider the transfer matrix with finite but small energy in the presence of nearest-neighbor hopping given in Eq.\eqref{eq:3}. For each step, the transfer matrix $M_i$ is given as\cite{kohmoto1987critical}, 
\begin{align}
\label{eq:21}
&M_i=\begin{pmatrix} \frac{E}{t_i} & -\frac{t_{i-1}}{t_i} \\ 1 & 0  \end{pmatrix}.
\end{align}
Then, the wave functions $\psi(n)$ and $\psi(n-1)$ is represented using the transfer matrix $M_i$, $\psi(0)$ and $\psi(1)$,
\begin{align}
\label{eq:22}
&\begin{pmatrix} \ \psi(n) \\ \psi(n-1)  \end{pmatrix}=\prod_{i=1}^{n-1} M_i \begin{pmatrix} \ \psi(1) \\ \psi(0)  \end{pmatrix}.
\end{align}
Here, $\psi(1)$=$\frac{E}{t_0}\psi(0)$ from Dirichlet condition mentioned above. With defining $\mathbb{M}(n)$=$\begin{pmatrix} \ m_{11} (n) & m_{12} (n) \\  m_{21}(n)  & m_{22}(n) \end{pmatrix}=\prod_{i=1}^{n-1} M_i$, one gets the following expression,
\begin{align}
\label{eq:23}
&\psi(n)=\left(m_{12}(n)+m_{11}(n)\frac{E}{t_0}\right)\psi(0).
\end{align}

Within perturbative regime near zero energy, the transmittance is obtained from Eq\eqref{eq:20} by collecting suitable orders of the energy correction. Here, we have assumed that energy (at least near zero energy) behaves like almost continuous variable.
This assumption is reasonable based on the gap labeling theorem and the shape of IDOS where \textit{countably many distinct tiny} gaps are present  around $E=0$ (a.k.a. singular continuous spectrum)\cite{makela2018mean,rai2019proximity,kellendonk2015mathematics}. Thus, considering the region around $E=0$ where only tiny size gaps are allowed from its singular continuous spectrum, one can assume the spectrum as a continuous variable near $E=0$\cite{kellendonk2015mathematics}. 

Having known the validity of perturbative approach near $E=0$, we now consider the generation of a plane wave at the zeroth position $i=0$  and compute the leading order contributions of $E$ in transfer matrix $\mathbb{M}(n)$ near zero energy. This corresponds to take into account the system with $n+1$ number of sites and $n$ number of links.
From simple algebra, it is easy to check that orders of energy correction in (off-) diagonal elements of $\mathbb{M}(n)$ depend on evenness and oddness of site $n$. For odd $n$, for instance, diagonal elements (off-diagonal elements) are expressed in terms of \textit{only} even (odd) orders of energy $E$, whereas it is reversed for even $n$. 

The first order energy correction of transmittance is obtained from the combination of the zeroth order of $E$ in diagonal element $m_{11}$ and the first order of $E$ in off-diagonal element $m_{12}$. 
As mentioned above, this corresponds to the case of odd $n=2k+1$, thus the total number of sites is $n+1$. Then, following the mathematical induction, one gets specific formula of matrix elements \textit{up to} the zeroth order and the first order respectively, which are given in Eq.\eqref{eq:24}.
\begin{eqnarray}
\label{eq:24}
m_{11}(n=2k+1)&\approx&(-1)^k\rho^{f(k)}, \\
m_{12}(n=2k+1)&\approx&(-1)^{k}\sum_{i=1}^k \rho^{h(i)}\frac{E}{t_{2i}}\rho^{f(k)-f(i)}. \nonumber
\end{eqnarray} 
To understand Eq.\eqref{eq:24}, let's think about our perturbative approach of transfer matrix $\mathbb{M}(n)$. First, the zeroth order correction for $m_{11}(n)$ is exactly derived from the same procedure discussed for Eq.\eqref{eq:7}, simply replacing $h(n)$ by $f(n)$.  For $m_{12}(n)$, one requires to compute the first order energy correction. From Eq.\eqref{eq:21}, it is trivial that this energy correction appear as a form of $\frac{E}{t}$. Thus, for $\mathbb{M}(n)=\prod_{j=1}^{n-1} M_j$, we choose a single site $j$ which gives the first order energy correction. Here, $j$ must be chosen to be an \textit{even} site because our tight binding Hamiltonian has the sublattice symmetry. 
For $l<j$, all products of $\rho^{a(l \to l+2)}$ for even $l$ contribute to $m_{12}$, thus it results in $\rho^{h(i)}$. Whereas, for $l>j$, all products of $\rho^{a(l \to l+2)}$ for odd $l$ contribute to $m_{12}$ and it gives rise to $\rho^{f(k)-f(i)}$.
 Similar methodology to get the transfer matrix elements can be used for higher order perturbation too. Again you can obtain this result from simple mathematical induction too as written in the Supplementary Materials Sec.\ref{subsec:boring calculation}. 
 For simplicity, we rewrite $m_{12} (n) +\frac{E}{t_0}m_{11} (n)$= $\rho^{h(1)}Eu(\rho,n,\mathcal{T})$ where $u(\rho,n.\mathcal{T})$ is defined as following.
\begin{align}
\label{eq:25}
&u(\rho,n=2k+1,\mathcal{T})=\sum_{i=0}^k \rho^{h(i)-h(1)}\frac{1}{t_{2i}}\rho^{f(k)-f(i)}.
\end{align}
Here, $f(0)$=$h(0)$=0 and $u(\rho,n\mathcal{T})$ is a function of $\rho$, $n$ and tiling $\mathcal{T}$ which determines both $h(i)$ and $f(i)$. 
Then, the transmittance for odd $n=2k+1$ is represented as following.
\begin{align}
\label{eq:26}
&T_{ n{(\textit{odd})}}(E) \approx \cosh^{-2}{(\kappa h(1)+\ln{\left | Eu(\rho,n,\mathcal{T}) \right |})}.
\end{align}
For odd $n$, it is clearly $T_{n}(E)\to0$ for $E\to0$. In this case, sublattice symmetry of the system guarantees absence of zero energy state at odd sites when the zero energy state is generated at even site $i=0$. Thus, it results in zero transmittance.

Exactly the same approach can be applicable for the second order correction in terms of energy $E$. 
In this case, one  should consider the first (second) order correction of $E$ in $m_{11}(n)$($m_{12}(n)$) and this corresponds to compute $\mathbb{M}(n)$ for even $n=2k$. Again based on the mathematical induction, it is easy to get the matrix elements up to the corresponding order of energy correction,  
\begin{widetext}
\begin{eqnarray}
\label{eq:27}
m_{11}(n=2k)&\approx&(-1)^{(k-1)}E\sum_{i=1}^k  \rho^{f(i-1)} \frac{1}{t_{2i-1}} \rho^{h(k)-h(i)}, \\
m_{12}(n=2k)&\approx& (-1)^k\rho^{h(k)}+(-1)^{(k-1)}E^2 \sum_{j=1}^{k-1}\sum_{i=j}^{k-1} \rho^{h(k-i)} \frac{1}{t_{n-2i}}\rho^{f(k-j)-f(k-i)} \frac{1}{t_{n-(2j-1)}} \rho^{h(k)-h(k-(j-1))}. \nonumber
\end{eqnarray}
\end{widetext}
Then, the transmittance for even $n=2k$ is given as following.
\begin{eqnarray}
\label{eq:28}
T_{n{(\textit{even})}}(E) \!\approx\!\cosh^{-2}(\kappa h(k) \!+\! \ln\left|1\!-\!E^2 w (\rho,n,\mathcal{T}) \right| ), \nonumber \\
\end{eqnarray}
where the pattern dependent function $w(\rho,n=2k,\mathcal{T})$ is defined as, 
\begin{eqnarray}
\label{eq:w-function}
w (\rho,&&n=2k,\mathcal{T})=\sum_{p=1}^k\frac{\rho^{f(p-1)\!-\!h(p)}}{t_{0}t_{2p-1}}   \\
&& \!+\! \sum_{j=1}^{k-1}\sum_{i=j}^{k-1} \frac{\rho^{h(k-i) \!+\! f(k-j)\!-\!f(k-i)\!+\!h(k)\!-\!h(k-(j-1))}}{t_{n-2i}t_{n-(2j-1)}}. \nonumber
\end{eqnarray}

\begin{figure}[!h]
\centering
\includegraphics[width=0.5 \textwidth]{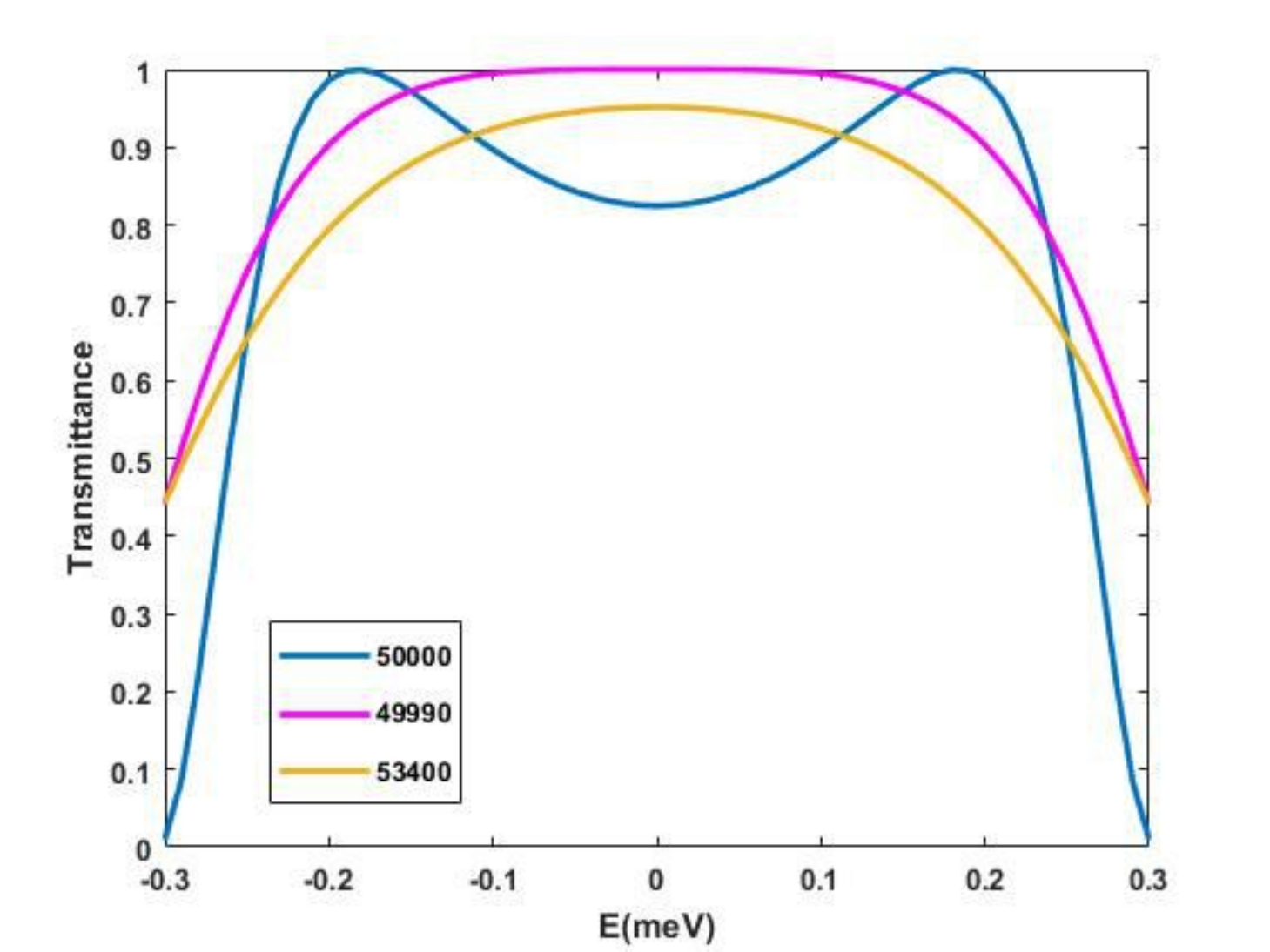}
\caption{\label{fig:4} $T_{n(even)}$($E$) up to the second order perturbation form, when the system is placed in air boundary. Concavity (or convexity) behavior of transmittance at $E=0$ is completely determined at the second order of $E$ in Taylor series of $T_{n(even)}(E)$. Depending on system size, the system shows convex, concave or neither. }
\end{figure}

Based on the analytic expression derived in Eq.\eqref{eq:28} with Eq.\eqref{eq:w-function}, 
Fig.\ref{fig:4} shows the transmittance for three different system sizes with even $n$, $T_{n {(\textit{even})}} (E)$,  as a function of energy $E$
for $\rho=0.9$ and $t_S=1$eV.

As shown in Fig.\ref{fig:4}, transmittance as a function of energy near $E=0$ strongly depends on the number of links $n$ (total number of sites $n+1$), showing convex or concave curvatures. For instance, the system with $n=50000$ shows convex, the system with $n=53400$ shows concave and the system with $n=49990$ shows neither convex nor concave near $E=0$. Concave (or convex) behavior of transmittance at $E=0$ can be fully understood by the second derivative of $T_{n(even)}(E)$ in Eq.\eqref{eq:28} with respect to the energy at $E=0$,
\begin{align}
\label{secondderivative}
&\left(\! \frac{d^2T_{n=2k} \!(E) \!}{dE^2}\! \right)_{\!E\!=\!0}\!  \! \!\!=\!4 w(\rho,\!2k,\! \mathcal{T}) \tanh\left(\kappa h({k})\right) \sech^2\left(\kappa h({k})\right)
\end{align} 
Clearly, $w(\rho,2k,\mathcal{T}) $ is positive from Eq.\eqref{eq:w-function} and hence, the concavity at $E=0$ is entirely determined by the sign of $\kappa h(k)$. For perfect periodic system, Eq.\eqref{secondderivative} becomes zero since $\kappa=0$. Hence, when the system is perfectly periodic, regardless of the system size, $T_{n=2k}(E)$ is neither concave nor convex function of $E$ near zero energy. In contrast, when the system is aperiodic, $\kappa h(k)$ is not a simple function as shown in Sec.\ref{sec:Topology}. In particular, for the case of the Fibonacci tiling, we set $\kappa<0$ i.e., $t_L<t_S$, and hence $T_{2k}(E)$ is a concave function at $E=0$ if and only if $h(l)>0$. Moreover, since the sign of $h(n)$ rapidly oscillates in the Fibonacci case, as seen in Fig.\ref{fig:2}, small changes in the system size enable to change the concave or convex behavior of the transmittance at $E=0$.

The concavity or convexity characteristic of transmittance also results in interesting conductance behavior. From the Landauer's formula in linear response theory\cite{cornean2005rigorous}, the conductance $G$ is represented as, 
\begin{align}
\label{eq:35}
&G=\frac{2e^2}{h}\int dE ~T_{n=2k}(E) \left( -\frac{\partial f_{\text{FD}}(E)}{\partial E} \right)
\end{align}
Here, $f_{\text{FD}}(E)$ is Fermi-Dirac distribution at energy $E$ and $T_{n=2k}(E)$ is transmittance for a given system $n=2k$ and energy $E$. 
Here, $-\frac{\partial f_{\text{FD}} (E)}{\partial E}$ plays a role of energy window i.e., \textit{meaningful} range of the integration for given temperature. 
At zero temperature, $G$ is nothing but  $\frac{2e^2}{h}T_{n=2k}(E_F)$, where $E_F=0$ is the Fermi energy of our system. For small but finite temperature, 
transmittance near $E=0$ is represented using the Taylor expansion with respect to the energy,
\begin{align}
\label{eq:taylor1}
&T_{2k}(E) = T_{2k}(E=0)+\xi E^2 +\mathcal{O}(E^4)
\end{align}
Here, $\xi$ is the Taylor expansion coefficient of the second order which corresponds to concave or convex behavior of transmittance at $E=0$ as we have discussed. We can numerically calculate the integration in Eq.\eqref{eq:35} with Taylor expansion in Eq.\eqref{eq:taylor1}. If we take the width of energy window as ${\sim }{20k_B T}$, then the conductance $G$ becomes
\begin{align}
\label{eq:rigorous}
&G~{\approx} ~\frac{2e^2}{h} T_{2k}(E=0)+\frac{6.558 e^2}{h} \xi (k_BT)^2 + \mathcal{O}(T^4)
\end{align}
Depending on the value of energy window, the specific value in front of $\xi$ above can be changed while being positive. 
Since the second term in Eq.\eqref{eq:rigorous} determines the leading order of temperature dependence of conductance in low temperature regime, it can be reversed according to the sign of $\xi$ i.e., concave or convex behavior of transmittance at $E=0$. For instance, positive $\xi$ (convex transmittance near $E=0$) implies that at low temperature regime conductance increases as temperature increases. 

Since the sign of $\xi$ is fully determined by $h(k)$ as shown in Eq.\eqref{secondderivative}, concave or convex characteristics of transmittance at $E=0$ and hence temperature dependence of conductance in low temperature regime are topologically robust phenomena. Moreover, with rapid sign changes of $h(k)$ and $f(k)$ in the Fibonacci tiling as shown in Fig.\ref{fig:2} (a), concavity of transmittance at $E=0$ and temperature dependence of conductance are very sensitively changing with the system size. Thus, accessible control of transmittance and conductance is possible via small changes of the system size, which is a unique feature of quasicrystals absent in any periodic lattice system.

\subsection{Transmittance with semi-infinite conducting leads near zero energy}
\label{subsec:Transmittance with two conducting leads}

Now, let's consider electronic transmittance of the Fibonacci quasicrystal when the quasi-periodic system is connected by conductors i.e., two semi-infinite periodic leads. In this case, there are critical difference compared to the system placed in air. 
Here, the energy is bounded due to the existence of leads in both sides and the periodic leads have finite hopping strength, thus it generates additional boundary conditions. 
In particular, the boundary condition at the zeroth position $\psi(0)$ is given by,
\begin{align}
\label{eq:29}
&t_0\psi(1)+t\psi(-1)=E\psi(0).
\end{align}
Here, $t$ is the hopping strength of semi-infinite periodic leads.
Within the periodic leads, let the factor $\lambda$ which satisfies 
 $\psi(i+1)=\lambda \psi(i)$ and then the Schr\"{o}dinger equation on the periodic leads satisfies,
\begin{align}
\label{eq:bloch-phase}
&t \lambda^2+t=E\lambda.
\end{align}
Eq.\eqref{eq:bloch-phase} gives two solutions, say $\lambda_{\pm}$ depending on the sign of imaginary part.
Clearly $\lambda_{+},\lambda_{-}$ are complex conjugate pair.
Now, let's assume that a plane wave coming from $-\infty$ and length of the quasi-periodic system is $n+1$. 
(The total number of sites indicated by yellow in Fig.\ref{fig:system} is $n+2$ from $i=0$ site to $i=n+1$ site.) 
Let $\psi(0)$ be $A+B$ where $A,B$ are amplitudes of the incoming and reflected wave respectively i.e., for all $i \le 0$, $\psi(i)=A\lambda_{+}^i+B\lambda_{-}^i$.
Then, using the boundary condition Eq.\eqref{eq:29} and $B=\psi(0)-A$, $A$ is expressed in terms of $\psi(1),\psi(0)$,
\begin{eqnarray}
\label{eq3}
A&=&\frac{t_0\psi(1)-t \lambda_-\psi(0)}{(E^2-4t^2)^{1/2}} .
\end{eqnarray}
On the other hand, at site $n+1$ (right end site of the Fibonacci chain), the transmitted wave only exists. Thus, it satisfies $\psi(n+2)=\lambda_+\psi(n+1)$ and the boundary condition yields,  
\begin{align}
\label{eq5}
&\psi(n)=\frac{t \lambda_-}{t_n}\psi(n+1).
\end{align}
So, we can get transmission coefficient $\tau=\frac{\psi(n+1)}{A}$ as following.
\begin{align}
\label{eq7}
&\tau=(E^2-4t^2)^{1/2}\frac{m_{11}\psi(1)+m_{12}\psi(0)}{t_0\psi(1)-t\lambda_-\psi(0)}.
\end{align}
Here, we used $m_{ij}$ abbreviating $(n+1)$ in $m_{ij} (n+1)$ for simplicity. Then, after some algebra, the transmittance at energy $E$ with the number of links $n+1$ for the Fibonacci quasicrystal is represented as, 
\begin{widetext}
\begin{eqnarray}
\label{eq:30}
T_{n+1}(E)=\frac{(4-(E/t)^2)}{\left[\left( m_{12} -\frac{t_n}{t_0}m_{21}  \right)+\frac{E}{2t} \left( \frac{t}{t_0}m_{11}-\frac{t_n}{t}m_{22} \right)\right]^2+\left( \frac{t}{t_0}m_{11}+\frac{t_n}{t}m_{22} \right)^2\left(1-\frac{E^2}{4t^2}\right)}.
\end{eqnarray}
\end{widetext}

Now, similar to the case of air boundary, let's consider perturbation of transfer matrix in terms of energy $E$ in $m_{ij}$.
For even $n=2k$, Eq.\eqref{eq:30} is rewritten by using matrix elements $m_{ij}$ in terms of $\rho$, $h(i)$ and $f(i)$,
\begin{eqnarray}
\label{eq:31}
T_{2k+1}(E)&=&\frac{(4-(E/t)^2)}{\left[a+\frac{E}{2t}b_{-}\right]^2+b_{+}^2\left(1-\frac{E^2}{4t^2}\right)},
\end{eqnarray}
with defining $a$ and $b_\pm$ as,
\begin{widetext}
\begin{eqnarray}
\label{eq:ab}
a~&=&2E\left[ \frac{1}{t_0}\rho^{h(k)}\cosh{\left(\ln{\left(\frac{t_n}{t_0}\right)}\right)}+\sum_{i=1}^{k-1} \frac{1}{t_{2i}} \cosh{\left(\left(f(k)-f(i)+h(i)\right)\kappa\right)} \right]+\mathcal{O}(E^3) \\
b_{+}&=&(\pm) \Bigg[ 2\cosh{ \left( f(k)\kappa-\ln\left(\frac{t_0}{t}\right)\right)}-E^2 ( \sum_{j=0}^{k-1}\sum_{i=0}^{k-j-1}\frac{t\rho^{f(k)+h(k-j)+f(k-j-i-1)}}{\rho^{f(k-j)+h(k-j-i)}t_0t_{n-2j}t_{n-2j-2i-1}} \nonumber \\
&&~~~~~ +\frac{t_n}{t}\sum_{j=0}^{k-2}\sum_{i=0}^{k-j-2} \frac{\rho^{h(k)+f(k-j-1)+h(k-j-i-1) }}{\rho^{h(k-j)+f(k-j-i-1)}t_{n-2j-1}t_{n-2j-2i-2}}) \Bigg]+\mathcal{O}(E^4) \nonumber \\
b_{-}&=&2\sinh{\left( f(k)\kappa-\ln\left(\frac{t_0}{t}\right)\right)}+ \mathcal{O}(E^2) \nonumber
\end{eqnarray}
\end{widetext}
Up to the second order of energy $E$ in denominator of transmittance, one can easily read off the transmittance derived in Eq.\eqref{eq:31} can be rewritten as, 
\begin{align}
\label{eq:33}
&T_{2k+1}(E) \approx \frac{4-(E/t)^2}{4\cosh^2{\left( f(k)\kappa-\ln\left(\frac{t_0}{t}\right) \right)}+\alpha E^2}, 
\end{align}
where $\alpha$ is given by
\begin{widetext}
\begin{eqnarray}
\label{terriblealpha}
&&\alpha\!=\! \left(\!2\!\left[ \!\frac{\rho^{h(k)}\! \cosh\!{\left( \! \ln \!{\left(\! \frac{t_n}{t_0}\! \right)}\!\! \right)}}{t_0}
\!+\! \sum_{i=1}^{k-1} \frac{\cosh\!{\left(\! \left(f(k)\!-\!f(i)\!+\!h(i) \! \right)\kappa \right)}}{t_{2i}} \! \right]\!+\!\frac{\sinh  {\left(\! f(k)\kappa \!-\! \ln\left(\! \frac{t_0}{t}\! \right) \! \right)}}{t} \!\! \right)^2\!\! 
\!-\! \frac{\cosh^2{\left(\!f(k)\kappa \!-\! \ln\left(\frac{t_0}{t}\right)\right)}}{t^2} \nonumber \\
&& \!-4\cosh\!{ \left(\! \! f(k)\kappa \!-\! \ln\! \left(\! \frac{t_0}{t}\! \right)\!\! \right)} \! \!\Bigg( 
\! \sum_{j\!=\!0}^{k\!-\!2} \! \sum_{i=0}^{k\!-\!j\!-\!2}\! \frac{t_n\rho^{h(k)+f(k-j-1)+h(k-j-i-1) }}{t\rho^{h(k-j)+f(k-j-i-1)}t_{n-2j-1}t_{n-2j-2i-2}} \!+ \! \sum_{j\!=\!0}^{k\!-\!1}\! \sum_{i\!=\!0}^{k\!-\!j\!-\!1}\! \frac{t\rho^{f(k)+h(k-j)+f(k-j-i-1)}}{\rho^{f(k-j)+h(k-j-i)}t_0t_{n-2j}t_{n-2j-2i-1}} \! \Bigg).
\nonumber \\
\end{eqnarray}
\end{widetext}
For odd $n=2k+1$, similar calculation can be performed and it gives no significant difference in transmittance, resulting in a similar form as Eq.\eqref{eq:33} which is derived for the case of even $n$. 

Here, one important concept is the \textit{half bandwidth} of transmittance, $\Delta_{1/2}$, the energy width of which transmittance becomes half of its maximum value i.e., $T_{n+1}\left(\frac{\Delta_{1/2}}{2}\right)=\frac{T_{n+1}(0)}{2}$. (Note that the critical point of Eq.\eqref{eq:33} uniquely exists at $E=0$, so the local extremum of transmittance is given at $E=0$.) Thus based on Eq.\eqref{eq:33}, one obtains,
\begin{align}
\label{eq:bandwidth}
&\Delta_{1/2}
=\frac{4t ~{\cosh\left(f(k)\kappa-\ln\left(\frac{t_0}{t}\right) \right)}}{\sqrt{2\cosh^2\left(f(k)\kappa-\ln\left(\frac{t_0}{t}\right) \right)+\alpha t^2}}.
\end{align}
It is important to note that the energy should be bounded by $|E|<2t$ due to the periodic leads and this further constrains $\Delta_{1/2}<4t$ to make Eq.\eqref{eq:bandwidth} a physically meaningful quantity. From Eq.\eqref{eq:bandwidth}, the constraint $\Delta_{1/2}<4t$ can be rewritten as following condition,
\begin{align}
\label{eq:bandwidthcondition}
&\alpha t^2+\cosh^2\left(f(k)\kappa-\ln\left(\frac{t_0}{t}\right) \right)>0,
\end{align}
and this is indeed equivalent to the condition of concave transmittance at $E=0$, $\left( \frac{d^2T_{2k+1}(E)}{dE^2} \right)_{E=0}<0$.

\begin{figure}[!h]
\centering
\includegraphics[width=0.5 \textwidth]{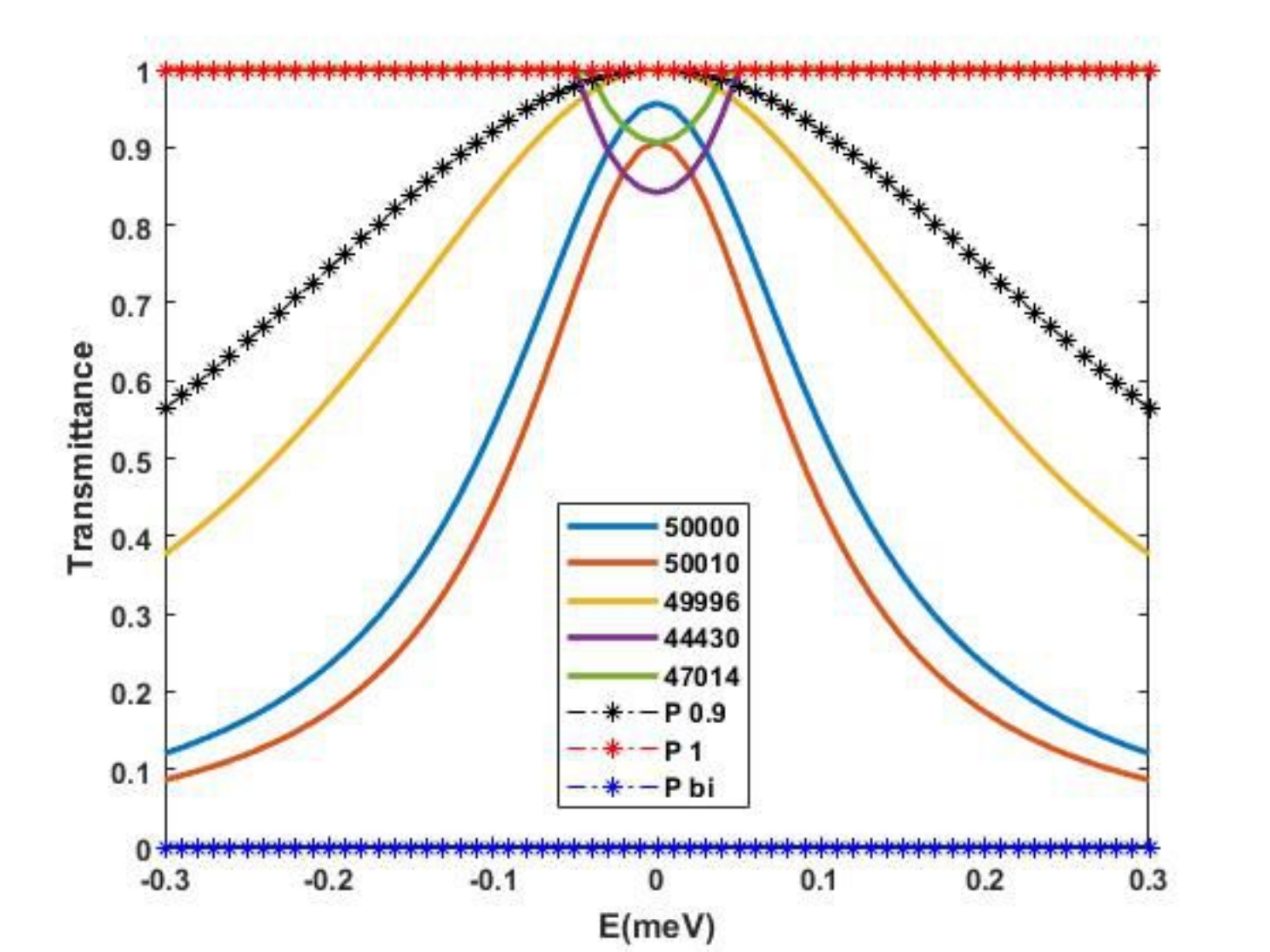}
\caption{\label{fig:5}
Transmittance as a function of energy $E$, $T_{2k+1}$($E$) when the system is placed in between two periodic conducting leads. (We set $\rho$=0.9 and $t_s=$1eV for quasicrystal and $t=1$ for conducting leads.) Concavity of transmittance at $E=0$ is a topologically protected characteristic and sensitively depends on the system size; concave behavior when $n=50000,50010$ and $49996$ and convex behavior when $n=44430$ and $47014$.  For comparison, periodic cases are also shown; a single prototile for hopping magnitudes $t=1$eV (P~1), $t=0.9$eV (P~0.9) and bi-prototile with alternating hopping magnitudes 1eV and 0.9eV (P~bi). See the main text for details.
} 
\end{figure}

Fig.\ref{fig:5} shows the transmittance near zero energy for different system sizes when the strength of quasi-periodicity is set to be $\rho=0.9$, $t_S=1$eV and the hopping strength on conducting leads is $t=1$ eV. 
For comparison, we also present three periodic cases with $n=50000$; a single prototile with hopping magnitudes 1eV (P1), 0.9eV (P 0.9)  and bi-prototile with alternating hopping magnitudes 1eV and 0.9eV (P bi).

We first discuss the transmittance for periodic cases which was already well understood. 
For a single prototile case with an equal magnitudes of hopping between system and leads  (See P$1$ in Fig.\ref{fig:5}.), the Bloch theorem guarantees perfect transmittance and this yields constant transmittance with a magnitude $1$. 
 In contrast, when the hopping magnitude in the system is smaller than the ones in connected leads (See P $0.9$ in Fig.\ref{fig:5}.), transmittance decays as energy $E$ deviates from zero. It is because the system is considered as  the disorderd region in the presence of leads.
To show concave transmittance regardless of the hopping strength say $t_P$ in the periodic system, it is enough to show that Eq.\eqref{eq:bandwidthcondition} holds for any $t_P$.
In the periodic limit, after some algebra, Eq.\eqref{terriblealpha} yields $\alpha=\left(\frac{n}{t_P} \right)^2 \left(\cosh^2{\left(\ln{\left(\frac{t_P}{t}\right)}\right)}-1\right)-\frac{1}{t^2}$. Thus, the left hand side of Eq.\eqref{eq:bandwidthcondition} is simplified as, 
\begin{align}
\label{eq:test}
&\left(\frac{nt}{t_P} \right)^2 \left(\cosh^2{\left(\ln{\left(\frac{t_P}{t}\right)}\right)}-1\right).
\end{align}
It is always positive regardless of $n$ and $t_P\neq t$, hence the condition Eq.\eqref{eq:bandwidthcondition} which is equivalent to concavity of transmittance holds and this yields
\begin{align}
\label{Bandwidthforperiodic}
&\Delta_{1/2}=\frac{4t}{\sqrt{1+\left(\frac{nt}{t_P} \right)^2\left(\cosh^2{\left(\ln{\left(\frac{t_P}{t}\right)}\right)}-1\right)}}
\end{align}
 Therefore, all periodic system with $t_P\neq t$ shows concave transmittance curve near $E=0$.
In addition, it is clear that smaller $t_{P}$ leads to smaller value of bandwidth $\Delta_{1/2}$.
 For bi-prototile case (See P bi in Fig.\ref{fig:5}.) shows zero transmittance guaranteed by sublattice symmetry of the Hamiltonian and the gap labeling group $\frac{1}{2}$ which indicates the existence of energy gap at $E=0$.

On the other hand, for the Fibonacci quasicrystal system, concave or convex behavior of transmittance at $E=0$ depends on the system    
and Eq.\eqref{eq:bandwidthcondition} may not work for certain system size. This is in contrast to the periodic case where Eq.\eqref{eq:bandwidthcondition} is always satisfies regardless of the system size.
As shown in Fig.\ref{fig:5}, for instance, Eq.\eqref{eq:bandwidthcondition}  is failed for $n=44430$ (violet) and $n=47014$(green) and convex transmittance behavior is shown near $E=0$. Whereas, for $n=50000$(skyblue), $n=50010$(orange) $n=49996$(yellow), Eq.\eqref{eq:bandwidthcondition} is satisfied hence concave transmittance is shown near $E=0$.
In Fig.\ref{fig:5}, we have set the maximum transmittance of the Fibonacci case as 1, which in general it can be larger than 1 due to an artifact of the perturbation up to the second order. 
However, such modification does not change a unique concavity feature in transmittance.

\begin{figure}[!h]
\centering
\subfloat[]{\includegraphics[width=0.5 \textwidth]{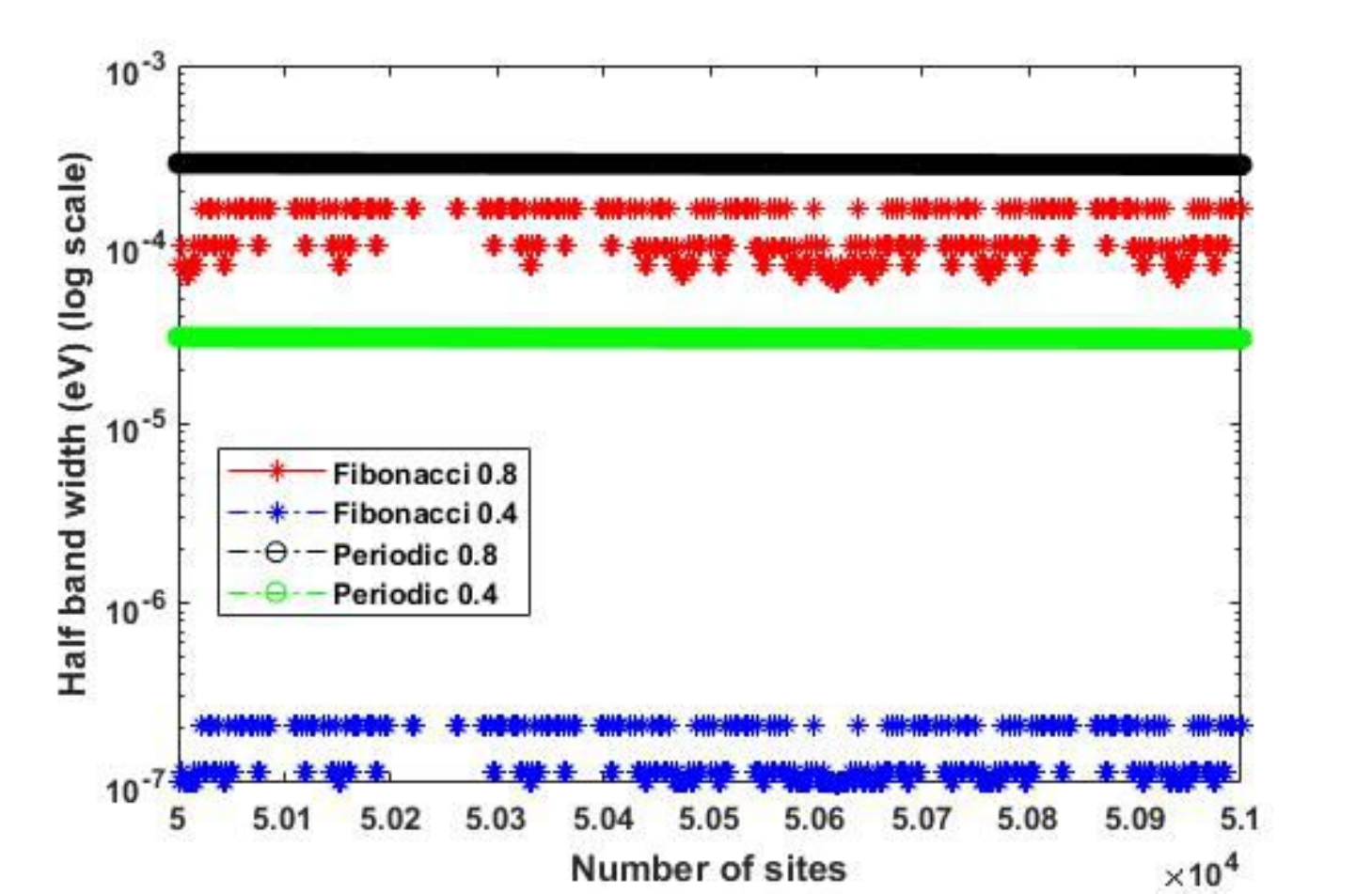}
\label{fig:6a}}
\hfill
\subfloat[]{\includegraphics[width=0.5 \textwidth]{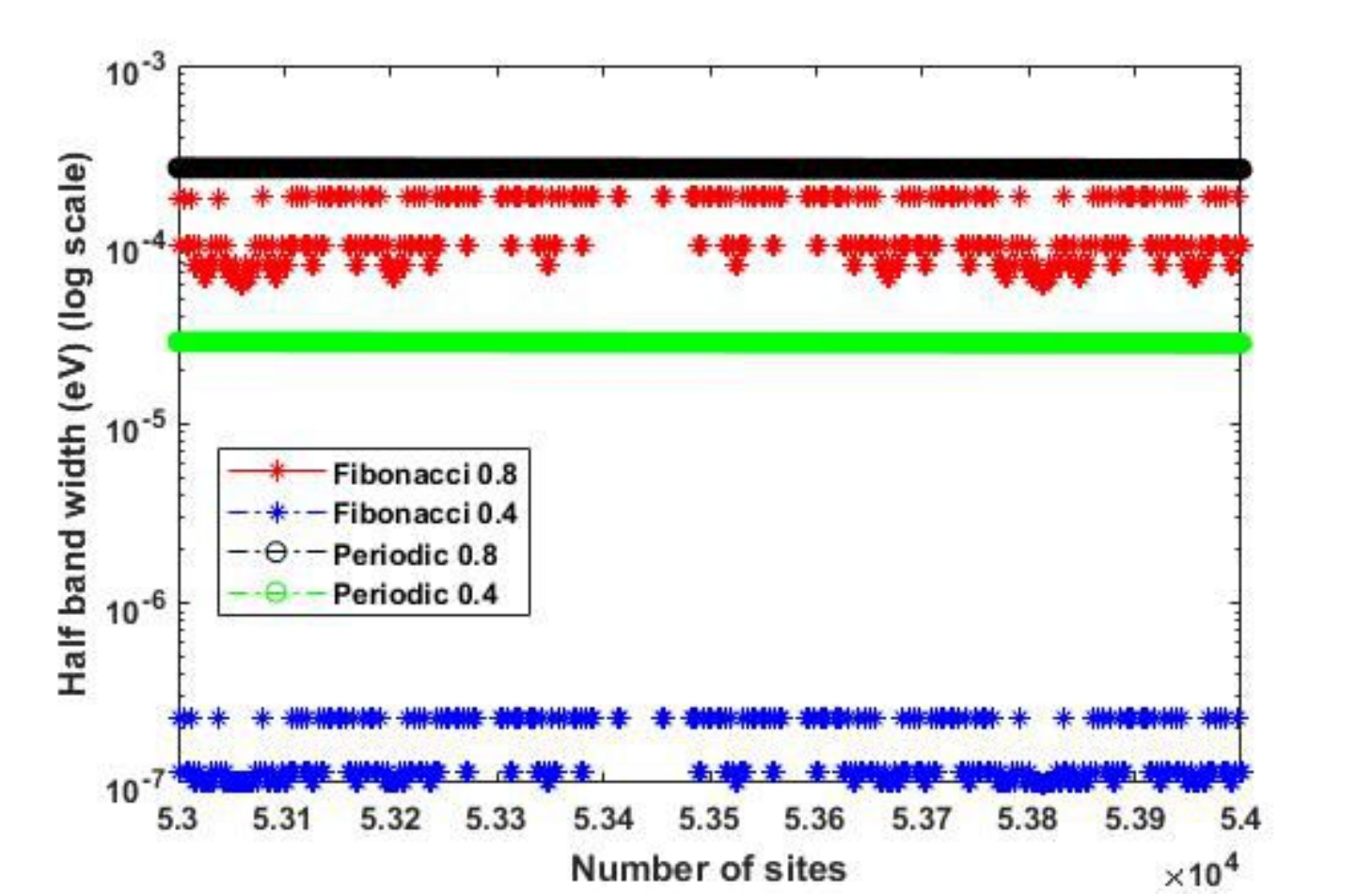}
\label{fig:6b}}
\caption{\label{fig:6} Log scale plot of half \textit{bandwidth} $\Delta_{1/2}$ as a function of number of sites in both Fibonacci quasicrystal and periodic case;  Fibonacci quasicrystal with $\rho=0.8$(red), $\rho=0.4$(blue) and periodic system $\frac{t_P}{t}=0.8$(black), $\frac{t_P}{t}=0.4$(green) respectively. (a) and (b) show $\Delta_{1/2}$ for different sections in total system especially in between $n=50000\sim51000$ and $n=53000\sim54000$ respectively.
Self-similar behavior of $\Delta_{1/2}$ for Fibonacci cases can be captured by comparing (a),(b) and the pattern itself does not depend on $\rho$ and $\frac{t_P}{t}$. See the main text for details.}
\end{figure}

To capture the existence and dependence of the half \textit{bandwidth} $\Delta_{1/2}$ defined in Eq.\eqref{eq:bandwidth}, Fig.\ref{fig:6} shows numerical results of $\Delta_{1/2}$ (if exists) in the Fibonacci quasicrystal for different strength of quasi-periodicity $\rho=0.4$ (Fibonacci 0.4) and $\rho=0.8$ (Fibonacci 0.8). In particular, Fig.\ref{fig:6a} and Fig.\ref{fig:6b} show $\Delta_{1/2}$ for different number of sites ranging from 50000 to 51000 and from 53000 to 54000 respectively. 
For comparison, $\Delta_{1/2}$ of the periodic limit is also presented for $\frac{t_P}{t}=0.4$ (Periodic 0.4) and $\frac{t_P}{t}=0.8$ (Periodic 0.8).
In Fig.\ref{fig:6}, both cases `Fibonacci 0.8' and `Fibonacci 0.4' show many number of sites where $\Delta_{1/2}$ does not exist, thus no red or blue marks are present. As explained earlier, this indicates convex character of transmittance near zero energy at these given sites. 
In addition, strength of quasi-periodicity also affects to the value of bandwidth. The bandwidth $\Delta_{1/2}$ becomes smaller for smaller $\rho$.  Hence regardless of tiling pattern, stronger quasi-periodicity makes sharper transmittance curve as long as Eq.\eqref{eq:bandwidthcondition} holds. %

In Fig.\ref{fig:6}, we note some important remarks. First of all, the pattern of $\Delta_{1/2}$ is independent of the strength of quasi-periodicity, $\rho$. By comparing `Fibonacci 0.8' and `Fibonacci 0.4',
one can see that the \textit{pattern} of red and blue marks for $\Delta_{1/2}$ are independent of $\rho$, although magnitude of $\Delta_{1/2}$ can be changed with respect to $\rho$.
 Second, the \textit{pattern} of $\Delta_{1/2}$ shows self-similar behavior which is captured by comparing Fig.\ref{fig:6}(a) and Fig.\ref{fig:6}(b). Such self-similarity originates from the fact that $h(k)$ and $f(k)$ in Eq.\eqref{eq:bandwidth} is self-similar functions as argued in Sec.\ref{sec:Topology}. Third, unlike the periodic case, small changes of the system size in Fibonacci tiling can sensitively control $\Delta_{1/2}$ 
and the concavity of transmittance can be also sensitively changed by small changes in number of sites.

Now let's discuss more details for concavity condition in transmittance that is identified by Eq.\eqref{eq:bandwidthcondition}. 
Failure of Eq.\eqref{eq:bandwidthcondition} originates from negative value of $\alpha$ and in thermodynamic limit, this value $\alpha$ is dominated by the last double sum term in Eq.\eqref{terriblealpha}. 
Since $\rho<1$ and the last double sum term is scaled by $\rho^{f(k)}$, negative value of $f(k)$ leads to a large negative value $\alpha$, resulting in failure of Eq.\eqref{eq:bandwidthcondition}. In contrast, positive $f(k)$ implies that small magnitude of the last double sum term in Eq.\eqref{terriblealpha}, so that Eq.\eqref{eq:bandwidthcondition} is valid. Unfortunately, when exponent of scaling factor $f(k)$ is zero, we cannot build up such rough estimation, instead we should concern additional information about tiling pattern such as $f(k-1)$ etc.
Hence, based on above argument, we cannot make any conclusion about concavity of transmittance for $f(k)=0$, i.e., \textit{undetermined}.
Above argument can be summarized as following,
\begin{align}
\label{eq:con}
&sgn\left(\left(\frac{d^2T_{2k+1}(E)}{dE^2}\right)_{E=0}\right)=\begin{cases} +1, f(k)< 0 \\ -1, f(k)>0 \\ undetermined, f(k)=0. \end{cases}
\end{align}
Here, $sgn(x)$ is the sign of $x$. Because $f(k)$ is a unique topological quantity in the Fibonacci quasicrystal, Eq.\eqref{eq:con} implies that even when the system is connected with semi-infinite periodic leads, concave or convex behavior of transmittance is topologically protected phenomena for non-zero $f(k)$. Furthermore, especially due to the rapidly oscillating behavior of $f(k)$ in Fibonacci tiling, Eq.\eqref{eq:con} implies that small change in number of sites enables to alter the concavity character of transmittance curve at $E=0$.
One can also numerically check if our argument holds and Eq.\eqref{eq:con} is valid.  
\begin{figure}[!h]
\centering
\includegraphics[width=0.5 \textwidth]{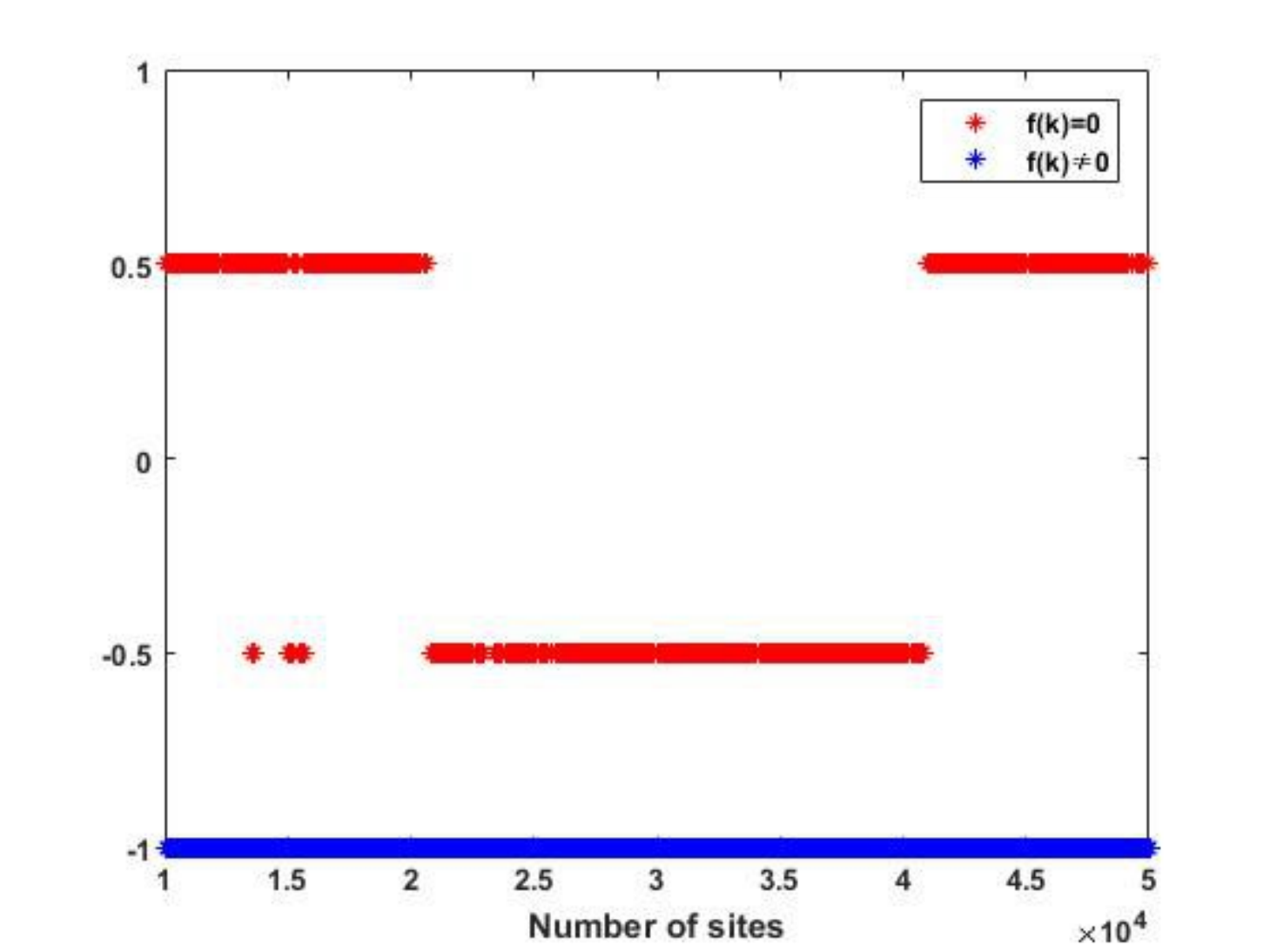}
\caption{\label{fig:bandfinal} Check validity of Eq.\eqref{eq:con} for the Fibonacci system with $\rho=0.99$, which relates between concave or convex characters of transmittance at zero energy and $f(k)$.
For the sites with $f(k)\!\neq\!0$, blue points show the plot of $sgn(f(k) \times \left({d^2T_{2k+1}(E)}/{dE^2}\right)_{E=0} ) $. It shows a constant -1. 
For the sites with $f(k) =0$, red points show the plot of $sgn(\left({d^2T_{2k+1}(E)}/{dE^2}\right)_{E=0} )/2$. In this case, concavity depends on number of sites, thus \textit{undetermined}.
}
\end{figure}
In Fig.\ref{fig:bandfinal}, blue points show the plot of $sgn\left( f(k) \times(\frac{d^2T_{2k+1}(E)}{dE^2})_{E=0}\right)$ for the sites with $f(k)\neq 0$ and red points show the plot of $sgn\left( (\frac{d^2T_{2k+1}(E)}{dE^2})_{E=0}\right)/2$ for the sites with $f(k)= 0$.
As indicated from Eq.\eqref{eq:con}, blue points for $f(k)\neq0$ show a constant -1, whereas, red points for $f(k)=0$ show $\pm0.5$ depending on sites, thus \textit{undetermined}. This implies that our conclusion Eq.\eqref{eq:bandwidthcondition} holds for both zero and non-zero $f(k)$ cases.

\section{Conclusion}
\label{sec:results}

\begin{figure}[!h]
\centering
\includegraphics[width=0.5 \textwidth]{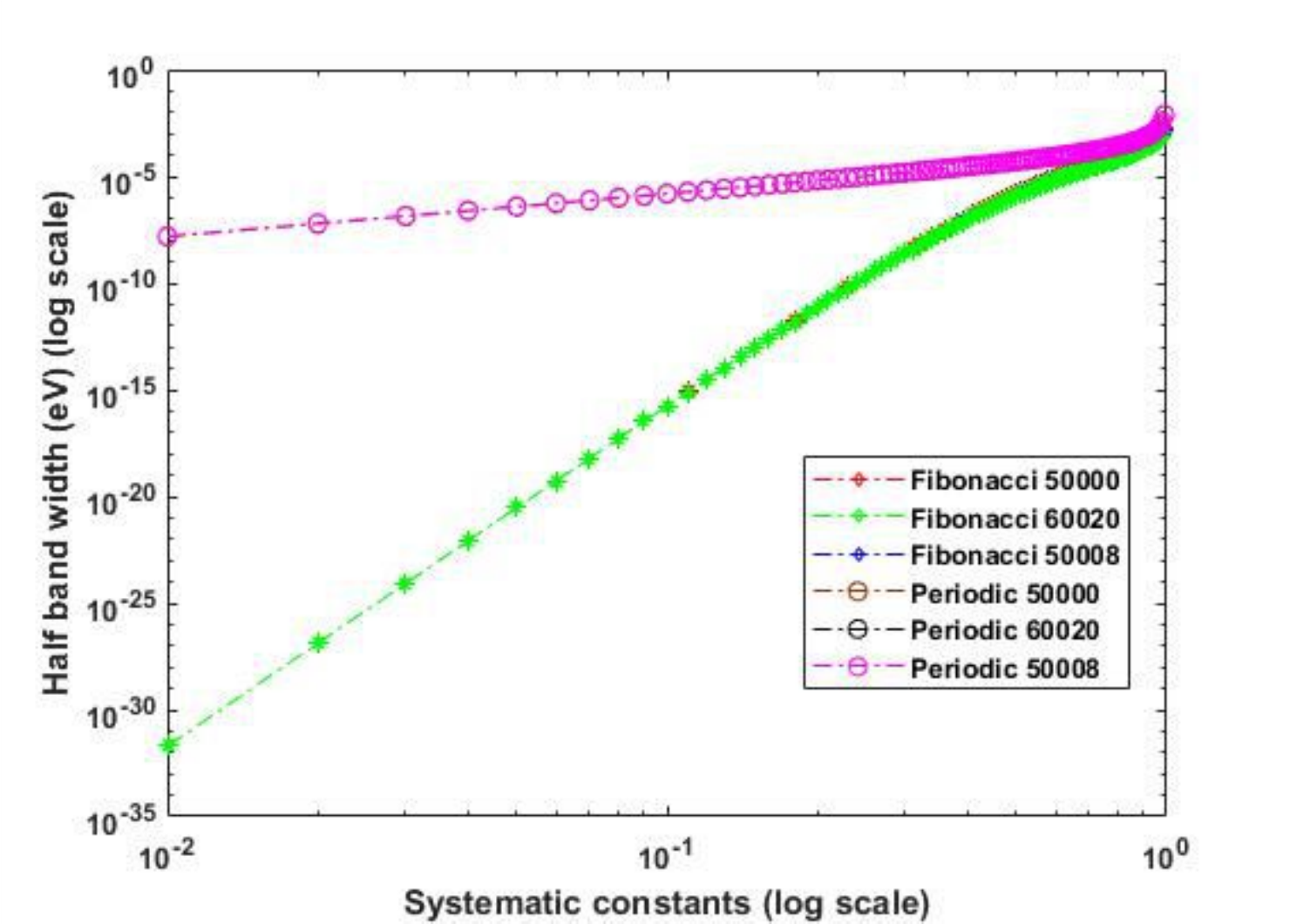}
\caption{\label{fig:7} Log-log plot of half-bandwidth of transmittance $\Delta_{1/2}$ as functions of systematic constants $\rho$ (for Fibonacci), ${t_P}/{t}$ (for periodic). Red, green and blue dots show $\Delta_{1/2}$ for the Fibonacci quasicrystal, whereas, brown, black and pink dots show $\Delta_{1/2}$ for periodic case, for the system size $n\!=\!50000$, $n\!=\!60020$ and $n\!=\!50008$ respectively.  For small $\rho$ and ${t_P}/{t} $, $\Delta_{1/2}$ does not depend on the system size but solely depends on systematic constants with distinct exponents between Fibonacci and periodic cases i.e., $\Delta_{1/2}\! \propto \! \rho^\gamma$ and $\Delta_{1/2} \!\propto \!(t_P/t)^\beta$ where $\gamma \!\neq \!\beta$. See the main text for details.}
\end{figure}

Before we conclude, let's discuss an interesting scaling behavior of half-bandwidth of transmittance as a function of quasi-periodic strength. When the Fibonacci quasicrystal is sandwiched by two semi-infinite conducting leads, 
 Fig.\ref{fig:7} represents the log-log plot of half-bandwidth $\Delta_{1/2}$ as a function of systematic constant $\rho$. For comparison, we also plot $\Delta_{1/2}$ as a function of $t_P/t$ for the periodic case. 
For the systems that exhibit concave character of transmittance, we consider three different system sizes;  $n=50000, 50008$ and $60020$ (red, green and blue dots for Fibonacci and brown, black and red for periodic). For both Fibonacci and periodic systems, $\Delta_{1/2}$ shows almost linear behavior with respect to the systematic constants $\rho$ and $\frac{t_P}{t}$, regardless of the system size. This indicates scaling behavior, for instance $\Delta_{1/2}\propto \rho^{\gamma}$ in strong quasi-periodic regime of the Fibonacci quasicrystal ($\rho \ll 1 $), with some exponent $\gamma$ that strongly depends on tiling pattern itself rather than the system size. On the other hand, the scaling behavior for the periodic case shows $\Delta_{1/2}\propto (t_P/t)^{\beta}$ for $(t_P/t) \ll1$ with distinct exponent $\beta \neq \gamma$.
Moreover, such exponent is larger for the Fibonacci system than the periodic system i.e., $\gamma > \beta$, thus one can conclude that transmittance is much sensitively changed by the systematic parameter in the Fibonacci quasicrystal compared to the periodic case.
Considering the expression of $\Delta_{1/2}$ and it's relation to PE cohomology, we expect that such exponent $\gamma$ might be not only strongly pattern dependent but also PE topologically robust quantity. As a future work, it is worth to show $\gamma$ as a topological quantity too and classify the aperiodic tilings based on $\gamma$ which implies the sensitivity of transmittance with respect to the quasi-periodicity.

In this paper, we have discussed pattern equivariant (PE) topological quantities in quasi-periodic systems, especially focusing on the first PE cochain and it's integration. Based on both gap labeling theorem and PE cohomology, we showed classification of  general aperiodic tilings in terms of localization characteristics of states, at which integrated density of states corresponds to $ \frac{1}{2}$. Furthermore, we investigated that such classification for aperiodic tiling system is related to the topologically robust quantity. Exemplifying one dimensional Fibonacci quasicrystal with nearest neighbor tight binding model, we performed PE cohomology group calculation especially focusing on a Barge-Diamond complex of supertilings. It turns out that the zero energy critical state is indeed topologically protected and it can give rise to unique scaling behavior. Furthermore, their topological properties can also induce non-trivial transmittance near zero energy. Based on the perturbative approach, we also explored electronic transmittance for two different cases (1) the system is placed in air (2) the system is connected by semi-infinite periodic leads. Unlike periodic case, transmittance of quasicrystal system turns out to be very sensitive to small changes in system size and strength of quasi-periodicity. In addition, we identified the concavity or convexity characters of transmittance near zero energy are completely determined by PE topological quantities. 
Here, we emphasize our new theoretical approach based on PE cohomology generally works for any aperiodic systems and it will give useful classification of localization characters of states.
As discussed in Sec.\ref{sec:Topology} exemplifying other aperiodic cases such as silver mean, Cantor set and binary non-pisot, the systems that the gap labeling theorem guarantee zero energy states i.e., gap labeling group does not include $1/2$, can be classified in terms of extended, critical and localized zero energy states. Thus, although there exist many distinct tilings, they may belong to the same PE cohomology group showing equivalent scaling behavior of  critical states.  
In addition, transmittance derived from a transfer matrix is a generic form which is applicable to both quasi-periodic and aperiodic systems.
Thus, the same argument for concavity or convexity characters of transmittance and their topological properties holds for different systems. This can further allows us to speculate robust transport behavior (e.g. low temperature conductance as functions of system size and strength of quasi-periodicity), which is originated from classification of general aperiodic systems. 
Our PE cohomology analysis for quasicrystals is not limited to the specific types of physical interactions. Thus, it is applicable not only to the electronic systems but also to the cases such as thermal system with phonon interaction, magnetic system with spin-spin exchange interactions and etc, which we leave as a future work \footnote{Junmo Jeon and SungBin Lee, in preparation.}.

\subsection*{Acknowledgement}
We are greateful to Hee Seung Kim for useful comments. This work is supported by the KAIST startup, BK21 and National Research Foundation Grant (NRF-2017R1A2B4008097).

\bibliography{my1}

\clearpage
\newpage

\section{Supplementary Materials}
\label{sec:supplement}

\subsection{Review: Scaling behavior of $h(n)$}
\label{subsec:Scaling behavior of h(n)}

Let's briefly review that the scaling behavior of $h(n)$ esepcially in the Fibonacci quasicrystal. This work is already well performed in previous studies\cite{mace2017critical,kohmoto1987critical}. In the Fibonacci quasicrystal, there are three types of length 2 supertiles, $LL,LS,SL$, say $A,B$ and $C$ respectively. We need the Fibonacci quasicrystal with total length to be even for well defined $h(n)$. 
Let's count the number of supertiles $A,B$ and $C$ for given $h(n)=h$. Since a single application of the substitution matrix leads the value of $h$ to be reversed and shifted by $\pm1$ or $0$, we can consider the distribution of $h$ for a specific super tile. 
\begin{align}
\label{eq:11}
&N^{t+1}_\mu(-h)=\sum_{h'=-1}^1M(h')_{\mu\nu}N^t_\nu(h+h').
\end{align}
Here, $N_{\mu}^t(h)$ is the number of supertile $\mu$ whose $h(n)=h$ after $t$ times of substitutions. $\mu,\nu \in A,B,C$ and the matrix element $M(h')_{\mu\nu}$ indicates the number of $\mu$ supertile which have $h(n)=h'$ from the substitution matrix $S$  for $\nu$ supertile. Explicitly we have,
\begin{eqnarray}
\label{eq:12}
M(-1)&=& \begin{pmatrix} 0 & 0 & 0 \\ 1 & 0 & 0 \\ 0 & 0 & 0 \end{pmatrix} \\
M(0)~&=& \begin{pmatrix} 0 & 0 & 0 \\ 1 & 1 & 2 \\ 1 & 1 & 0 \end{pmatrix} \nonumber \\
M(1)~&=& \begin{pmatrix} 1 & 1 & 1 \\ 0 & 0 & 0 \\ 1 & 1 & 1\end{pmatrix}. \nonumber
\end{eqnarray}

Note that Eq.\eqref{eq:11} is a \textit{Fokker-Plank like equation}, as considering $t$ as time, $h$ as spatial variable and $M(\delta h)$ as transition matrix.\cite{mace2017critical} Thus, one can easily expect that its probability distribution is given by the normal distribution for large $t$. To know the explicit form of the distribution of $h(n)$, we may define the partition function for $h(n)$ as following.
\begin{align}
\label{eq:13}
&Z^{t}_{\mu}(\beta)=\sum_{h\in\mathbb{Z}}N^{t}_{\mu}(h)\exp(\beta h).
\end{align}
Using Eq.\eqref{eq:11}, we can obtain
\begin{align}
\label{eq:14}
&{\bold Z}^{t+2}(\beta)=K(-\beta)K(\beta) {\bold Z}^t(\beta)
\end{align}
where ${\bold Z}^t (\beta) = (Z^t_A(\beta), Z^t_B(\beta),Z^t_C(\beta))^T$ and  $K(\beta)$ is the weighted matrix of $M(h)$ defined as,
\begin{align}
&K(\beta)=\sum_{h'=-1,0,1}M(h')\exp(-\beta h').
\end{align}
Here, $\beta$ plays a role of the weight factor that is given by a function of quasi periodicity strength, for instance $\rho=t_L/t_S$ in the tight binding model. In $t\to\infty$ limit (i.e. thermodynamic limit), it is already well known that $Z^{2t}_{\mu}(\beta)\thicksim\lambda^t(\beta)f_{\mu}(\beta)$. Here, $\lambda(\beta)$ is the \textit{largest} eigenvalue of $K(-\beta)K(\beta)$ and $f(\beta)$ is the corresponding eigenvector. In our case, one can obtain $\lambda(\beta)$ from a direct computation and the result is the following.
\begin{align}
\label{eq:15}
&\lambda(\beta)=\left( \frac{(1+e^\beta)^2+\sqrt{(1+e^\beta)^4+4e^{2\beta}}}{2e^{\beta}} \right)^2.
\end{align}
Thus, asymptotic distribution of $h(n)$ in thermodynamic limit is given by normal distribution as Eq.\eqref{eq:16} below.
\begin{align}
\label{eq:16}
&P_{\mu}(h)\thicksim\frac{f_{\mu}}{\sqrt{4{\pi}Dt}}\exp\left(-\frac{h^2}{4Dt}\right)
\end{align}
Here, $D$ is a diffusion coefficient of Fokker-Plank like equation\cite{9780444529657}, and hence $D=\frac{1}{4}\left(\frac{\partial^2 \ln(\lambda(\beta))}{\partial \beta^2}\right)_{\beta=0}$ in thermodynamic limit.

\begin{figure}[!h]
\centering
\includegraphics[width=0.5 \textwidth]{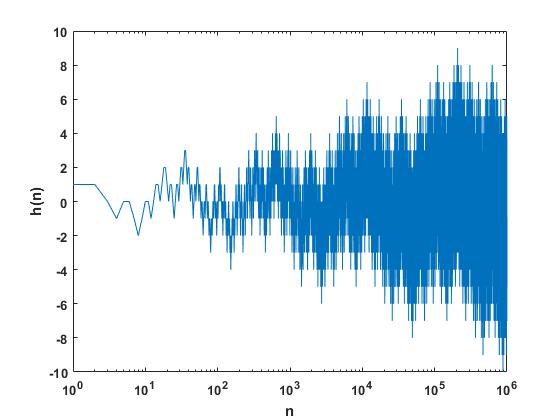}
\caption{\label{fig:11}$h(n)$ of Fibonacci quasicrystal up to $n=10^6$ (log scale). The oscillating amplitude of $h(n)$ is growing as function of $n$ extremely slowly (as Eq.\eqref{eq:hcomplex}) compare to Cantor set or binary non-Pisot cases.}
\end{figure}

The scaling behavior of $h(n)$ is given by the standard deviation of Eq.\eqref{eq:16} i.e. $h^2\thicksim2Dt$. As discussed in the main text, one of the eigenvalues (a.k.a. PV eigenvalue) of the substitution matrix is the golden ratio, $\tau$,  and hence, the system size is being $L_t\thicksim\tau^{3t}L_0$ in thermodynamic limit where $L_0$ is the length of the $S$  prototile. In small $\beta$ limit i.e., for weak quasi-periodicity $\rho \approx 1$, we may approximate $\ln(\lambda(\beta))= \ln (\lambda(0)) + 2D \beta^2 + \mathcal{O}( \beta^4)$. Based on above, we get the scaling behavior of $h(L)$ as a function of the system size $L$ as following. 
\begin{align}
\label{eq:hcomplex}
&h(L)^2\thicksim \frac{1}{3\beta^2 \ln \tau} \ln\left(\frac{\lambda(\beta)}{\lambda(0)}\right) \ln\left(\frac{L}{L_0}\right).
\end{align}
Here, the scaling behavior of $h(L)\thicksim \sqrt{\ln\left(\frac{L}{L_0}\right)}$ is significant, which is one of the unique features of the critical state\cite{mace2017critical}. 
In Fig.\ref{fig:11}, $h(n)$ is shown up to the system size $n=10^6$ and it indeed exhibits the logarithmic scaling behavior of Eq.\eqref{eq:hcomplex}.

\subsection{Supertiling cohomology groups of 1D tilings other than Fibonacci}
\label{subsec:silvermean}
In this section, we derive the first cohomology group of silver mean, Cantor set and binary non-Pisot systems. Let's remind that their substitution rules of $A,B$ prototiles.
\begin{eqnarray}
\sigma_{SM}~~&=&\begin{cases}A \to BAA \\ B \to A  \end{cases}  \\
\sigma_{CS}~~&=&\begin{cases}A \to ABA \\ B \to BBB  \end{cases} \nonumber \\
\sigma_{B-NP}&=&\begin{cases}A \to AB \\ B \to AAAAA \end{cases}. \nonumber
\end{eqnarray}
Here, $\sigma_{SM},\sigma_{CS},\sigma_{B-NP}$ are substitution rules of silver mean, Cantor set and binary non-Pisot system respectively.

Let's first consider the case of silver mean tiling. By successively applying $\sigma_{SM}$ to prototile $A$, we can get substitution generations as $A$, $BAA$, $ABAABAA$, $BAAABAAABAAABAABAA$ etc. So, it is easy to note that if one reads them from left to right, they can be grouped as even and odd generations depending on starting prototile. In contrast, if one reads it from right to left, all generations are unified. It implies that silver mean tiling is non-palindromic like Fibonacci. Note that breaking directions are oppositie between Fibonacci and silver mean tilings.
On the other hand, from above substitution generations, it is clear that there are only three kinds of 2-supertiles, say $AA\overset{\underset{\mathrm{def}}{}}{=}X$, $BA\overset{\underset{\mathrm{def}}{}}{=}Y$ and $AB\overset{\underset{\mathrm{def}}{}}{=}Z$. There is no $BB$ supertile again. If we consider the transmittance of the system from left to the right like Fibonacci, we collect substitution generations of either even or odd ones only. Hence, we consider \textit{twice} substitution on supertiling. It can be easily shown that the result of the cohomology is not changed when we consider the reverse the direction i.e., from right to left. By directly applying $\sigma_{SM}$, we can obtain that
\begin{eqnarray}
\label{silversupsub}
X &\to& ZXYXYZX  \\
Y &\to& YXYZX \nonumber \\
Z &\to& ZXYZX \nonumber
\end{eqnarray}
Therefore, the corresponding substituion matrix is given by (where ordered basis is chosen as $\left\{ X,Y,Z \right\}$) following.
\begin{align}
\label{eq:Smatrixofsilvermean}
&S=\begin{pmatrix} 3 & 2 & 2 \\ 2 & 2 & 1 \\ 2 & 1 & 2 \end{pmatrix}
\end{align}
The eigenvalues of this matrix are $1,3\pm 2\sqrt{2}$. Each of them implies that identity, \textit{irrational} inflation, \textit{irrational} contraction respectively. Now again from simple consideration of substitution rule, it is clear that there are only five kinds of vertex flips, $v_{XY}$, $v_{XZ}$, $v_{YX}$, $v_{YZ}$, $v_{ZX}$. Furthermore, their eventual range of substitution map is $\left\{v_{XY}, v_{XZ} \right\}$.

Hence, it is exactly the same as the Fibonacci quasicrystal case. If one define  inverse limit of Barge-Diamond complex (a.k.a. BD complex) as $\Xi$, and of vertex flips as $\Xi_0$, then BD complex and its quotient by vertex flips are both connected so that the reduced cohomologies $\check{H}^0(\Xi)=\check{H}^0(\Xi,\Xi_0)=0$. On the other hand, $\check{H}^1(\Xi,\Xi_0)=\mathbb{Z}^3$ which is the direct limit of $S^T$. In addition, eventual range of vertex flips is connected ($\check{H}^0(\Xi_0)=0$) and no loop ($\check{H}^1(\Xi_0)=0$). Finally from the long exact sequence for reduced cohomologies, we get $\check{H}^1(\Xi)=\mathbb{Z}^3$. This is exactly the same cohomology group with the Fibonacci quasicrystal, thus we can conclude that the Fibonacci and silver mean tilings have common topological behavior of $h(n)$ and share the same behavior of critical zero energy states.

In contrast, the Cantor set is very different from both Fibonacci and silver mean tilings even though there are still three kinds of supertiles, say $BB\overset{\underset{\mathrm{def}}{}}{=}X'$, $AB\overset{\underset{\mathrm{def}}{}}{=}Y'$ and $BA\overset{\underset{\mathrm{def}}{}}{=}Z'$. First of all, it is easy to check that this Cantor tiling is palindromic that is whenever we read it from left or right, all substitution generations are unified. Hence we don't need to apply substitution rule in multiple times when we consider that the direct limit. In the orderd basis as $\left\{ X',Y',Z' \right\}$, we get the substitution matrix $S'$ for 2-supertiles in Cantor tiling.
\begin{align}
\label{eq:Smatrixofcantor}
&S'=\begin{pmatrix} 3 & 1 & 1 \\ 0 & 2 & 0 \\ 0 & 0 & 2 \end{pmatrix}
\end{align}
Its eigenvalues are $3,2,2$ and eventual range of substitution map on vertex flips is $\left\{v_{X'X'}, v_{X'Y'} v_{Z'X'} \right\}$ which is single connected component without loop.

The difference in a Cantor set case compared to the case of Fibonacci and silver mean, is the direct limit of the wedge of three circles via substitution procedure, i.e., $\check{H}^1(\Xi',\Xi_0')$ where $\Xi',\Xi_0'$ are inverse limits of BD complex and subcomplex (vertex flips) of the Cantor set respectively. In this case our eigenvalues are all non unity integers, and hence $\check{H}^1(\Xi')=\check{H}^1(\Xi',\Xi_0')=\mathbb{Z}[1/3] \oplus \mathbb{Z}[1/2]^2$. Here $\mathbb{Z}[1/n]$ is the set of elements $\frac{k}{n^m}$ where $k,m$ are integers.
The binary non-Pisot case is similar to the case of a Cantor set. Note that it is non-palindromic and using exactly the same method as before, one can easily show that its first cohomology group is $\mathbb{Z}[1/5] \oplus \mathbb{Z}^2$. Furthermore, by studying the eigenvectors of the substitution matrix, one can easily conclude that $h(n)$ belongs to $\mathbb{Z}[1/2]$ sector for a Cantor set and $\mathbb{Z}[1/5]$ sector for a binary non-Pisot which are different with $\mathbb{Z}$ sector in the case of Fibonacci and silver mean. Hence, we can conclude that a Cantor set and a binary non-Pisot show different \textit{topologically protected} behavior compare to Fibonacci and silver mean, exhibiting robust localized zero energy eigenstate instead of critical state.

\subsection{Derivation of transfer matrix elements in perturbative form}
\label{subsec:boring calculation}
Let's derive the transfer matrix elements of $\mathbb{M}(n)\!=\!\prod_{i=1}^{n-1} M_i\!=\!\begin{pmatrix} \ m_{11} (n) & m_{12}(n) \\  m_{21}(n)  & m_{22}(n)\end{pmatrix}$ in a perturbative form Eq.\eqref{eq:24} up to the first order of energy, using mathematical induction.
One can obtain Eq.\eqref{eq:27} based on the same method, so it is sufficient to derive Eq.\eqref{eq:24} only.
First of all, successive matrix multiplication of transfer matrices yields,
\begin{eqnarray}
\label{eq:transfer2product}
M_{n-1}M_{n-2}&=&\begin{pmatrix} \frac{E}{t_{n-1}} & -\frac{t_{n-2}}{t_{n-1}} \\ 1 & 0  \end{pmatrix}\begin{pmatrix} \frac{E}{t_{n-2}} & -\frac{t_{n-3}}{t_{n-2}} \\ 1 & 0  \end{pmatrix}\\
&\approx& \begin{pmatrix} -\frac{t_{n-2}}{t_{n-1}} & -\frac{t_{n-3}}{t_{n-2}}\frac{E}{t_{n-1}} \\ \frac{E}{t_{n-2}} & -\frac{t_{n-3}}{t_{n-2}}  \end{pmatrix}. \nonumber
\end{eqnarray}
Here,  the second order of energy term in  the (1,1) matrix element has been neglected in the final result of Eq.\eqref{eq:transfer2product}, because we are considering first order perturbation now. The (1,1) matrix element is zeroth order of energy (i.e. constant), the (1,2) matrix element is linear function of energy. In order to apply mathematical induction, we also need to consider the form of (2,1) and  (2,2) matrix elements too. Suppose we multiply even number of transfer matrices, then  (2,1) matrix element contains odd order of $E$ only, (i.e. in a perturbative form it is a linear function of energy) and (2,2) element is given by $(-1)^k\rho^{h(k)}$ in perturbative form. Now we apply mathematical induction for entire four transfer matrix elements.
First, one can easily check that it clearly holds for the case of $k=1$ i.e. $n=3$ using Eq.\eqref{eq:transfer2product}. Next let's assume that our proposition above holds for the case of $n=2k+1$.
As a result, the direct computation yields for $n+2=2k+3$, 
\begin{widetext}
\begin{eqnarray}
\label{eq:transfer2product2}
\mathbb{M}(n+2)&\approx&\begin{pmatrix} -\frac{t_{n}}{t_{n+1}} & -\frac{t_{n-1}}{t_{n}}\frac{E}{t_{n+1}} \\ \frac{E}{t_{n}} & -\frac{t_{n-1}}{t_{n}}  \end{pmatrix}\begin{pmatrix} (-1)^k\rho^{f(k)} & (-1)^{k}\sum_{i=1}^k \frac{\rho^{h(i)+f(k)-f(i)}E}{t_{2i}}  \\ cE+\mathcal{O}(E^3) & (-1)^k\rho^{h(k)}  \end{pmatrix} \\
&\approx& \begin{pmatrix} (-1)^{k+1}\rho^{f(k+1)} & (-1)^{k+1}\sum_{i=1}^{k+1} \frac{\rho^{h(i)+f(k+1)-f(i)}E}{t_{2i}}  \\ c'E+\mathcal{O}(E^3) & (-1)^{k+1}\rho^{h(k+1)}  \end{pmatrix}, \nonumber
\end{eqnarray}
\end{widetext}
where $c,c'$ are some coefficients. 
Eq.\eqref{eq:transfer2product2} shows that even for $n+2=2k+3$ our proposition for four transfer matrix elements still holds and hence our proof is done by mathematical induction.
\end{document}